\documentclass[12pt]{article}

\usepackage{newtxtext}
\usepackage{float}
\usepackage[varvw]{newtxmath}
\usepackage{graphicx}

\usepackage{braket}
\usepackage{multirow}
\usepackage{graphicx} 
\usepackage{array}
\usepackage{amsmath}
\usepackage[normalem]{ulem}

\usepackage[letterpaper,margin=1in]{geometry}

\renewenvironment{abstract}
	{\quotation}
	{\endquotation}

\date{}

\makeatletter
\renewcommand{\fnum@figure}{\textbf{Figure \thefigure}}
\renewcommand{\fnum@table}{\textbf{Table \thetable}}
\makeatother

\usepackage{scicite}
\usepackage{url}
\usepackage{hyperref}

\def\scititle{
Tracking and distinguishing slime mold solutions to traveling \textcolor{black}{salesperson} problems through synchronized amplification in the non-equilibrium steady state
}
\title{\bfseries \boldmath \scititle}

\author{
	Suyash Bajpai$^{\dagger}$,
    Masashi Aono$^{\ddagger}$,
	Philip Kurian$^{\ast \dagger}$ \and
	\textcolor{black}{\small$^{\dagger}$Quantum Biology Laboratory, Howard University, 2041 Georgia Avenue NW, Washington, DC 20060} \and
    \textcolor{black}{
    \small$^{\ddagger}$Amoeba Energy Co., Ltd., 2-15-1 Konan, Minato, Tokyo, Japan 108-6022} \and
	\small$^\ast$Corresponding author. Email: pkurian@howard.edu\and
}

\begin{document}

\maketitle

\begin{abstract}
The plasmodium of the true slime mold \textit{\textit{Physarum} polycephalum}—an ancient, unicellular, aneural organism—serves as a platform for studying the information-processing capacities of active matter. Previous experiments used \textit{Physarum}'s intricate morphological dynamics and photoavoidance in stellate chips to solve $N$-city traveling salesperson problems (TSPs) of up to eight cities, scaling linearly in time with TSP size. Optical feedback controlled by a modified Hopfield network illuminated specific lanes at regular intervals, prompting \textit{Physarum} to elongate or retract selected branches. When the illumination pattern stabilized in a non-equilibrium steady state, branches bifurcated reproducibly into solution and non-solution groups, with the former exhibiting lower-frequency, higher-amplitude, and more synchronized oscillations than the latter across 41 trials with valid TSP solutions. 
\textcolor{black}{\textit{Physarum}'s synchronization dynamics efficiently predict 100\% of selected solutions by the midpoint of the optical-feedback interval, achieving statistically significant (paired $t$-test, $p<0.005$) discrimination from alternate tours well before the non-equilibrium steady state. Observed frequency downconversions and synchronized power amplifications scale linearly and quadratically, respectively, for small-to-moderate TSP size, as captured by a toy model of energy redistribution with saturating optical absorption.} Tuning these features in native biomolecular chromophore networks may thus improve both the quality and efficiency of TSP solutions from \textit{Physarum}-based biocomputers, which exploit the effects of organismal-scale coherence.
\end{abstract}



\section{\label{sec:level1}Introduction}

Despite the lack of a centralized control unit, \textit{\textit{Physarum} polycephalum} possesses sophisticated information-processing capacities. If food sources are arranged in a given spatial configuration, it can deform to an optimal shape, connecting them and maximizing the nutrient absorption \cite{nakagaki2000maze,nakagaki2007minimum, tero2010rules, dussutour2010amoeboid, jones2014computation}. Similarly, it forms regular graphs \cite{baumgarten2010plasmodial}, can recapitulate a periodic stimulus when triggered at much later times \cite{saigusa2008amoebae}, and has been shown to solve a constraint satisfaction problem \cite{aono2007amoeba}. Such remarkable capabilities arise from the multiscale oscillatory dynamics of \textit{Physarum}, which are indicative of the organism's ability to process environmental information and solve challenging optimization problems, including the well-known traveling salesperson problem (TSP) \cite{aono2009amoeba, zhu2013amoeba, iwayama2016decision, zhu2018remarkable}.

The TSP is stated as follows: In a map of $N$ cities, find the shortest tour visiting each city once and return to the starting city. It is one of the fundamental optimization challenges with diverse applications in logistics, network design, and computational theory. In contrast to traditional silicon-based approaches, which are limited to solving only predefined problems, a \textit{Physarum}-based computer has demonstrated \textit{meta-problem solving}—the simultaneous search for both an unknown problem and its solution—in a four-city TSP setup~\cite{aono2010amoeba}. Here, the task was not merely to solve a given TSP instance but to dynamically construct the instance itself. This was accomplished using a stellate chip in which distinct sets of lanes represented problem-searching and solution-searching neurons (branches). The distance matrix was continuously updated based on the state of the problem-searching neurons. Once the system stabilized, \textit{Physarum} converged on a specific problem-solution pair. After convergence, the system destabilized and then re-stabilized to identify a new pair. Through this process, the system iteratively generated consistent problem-solution pairs. This emergent ability to simultaneously identify and solve problems stands in stark contrast to silicon-based systems, which require fully defined problem inputs and generally lack the capacity to autonomously explore or generate novel problem configurations. 

Many real-world optimization problems are formulated as variations of the TSP, with additional constraints and objectives. Examples include computer wiring, vehicle routing, robotic control, circuit board drilling, crystallography, and chronological sequencing \cite{1570854176485676800}. In light of these potential applications, it is desirable to solve instances of the TSP, like other NP-hard problems, in reasonable time.

Unlike conventional silicon-based systems, which operate sequentially or rely on parallelized algorithms, \textit{Physarum} computes through globally connected, highly redundant, and parallel processes—an inherently different computing paradigm characteristic of many living systems. Instead of executing strictly rule-based algorithmic operations, its computational optimization emerges dynamically through continuous interaction with the environment. This unique mechanism may explain why a \textit{Physarum}-based computer was observed to solve small yet non-trivial TSP instances in approximately linear time \cite{zhu2018remarkable}---a scaling behavior that remains highly ambitious for conventional algorithms, even for relatively small problem sizes.

Such remarkable problem-solving capabilities in \textit{Physarum}, particularly in the TSP, motivated us to explore what makes \textit{Physarum}-selected solutions unique. We aim to understand how \textit{Physarum} discerns and selects a high-quality TSP tour from all possible routes. This investigation forms the core of our study, where we analyze the amplitude, power spectrum, frequency, and phase components of \textit{Physarum}'s TSP solutions to uncover the key physical quantities -- beyond the external light stimuli -- that drive its decision-making behavior. We hypothesized that the synchronization index ($S$) of phases in \textit{Physarum} branches serves as a key metric in its solution-selection process. 

The shape-changing dynamics of \textit{Physarum} in a stellate chip have previously been exploited to solve combinatorial optimization problems \cite{aono2007amoeba,aono2009amoeba,zhu2013amoeba,iwayama2016decision,zhu2018remarkable} by incorporating a modified Hopfield ``neural" network for optical feedback (Fig. 1). Typically, \textit{Physarum} grows in all the lanes of the chip to maximize the absorption of nutrients. However, it retreats when exposed to broadband white light, demonstrating clear photoavoidance behavior at different wavelengths \cite{ueda1988action}. Although \textit{Physarum} changes shape during these TSP experiments, a volume increase (decrease) in one lane is essentially conserved by a decrease (increase) in the other lanes, assuming we hold the thickness of the central processing disk constant. It has been shown that this approximate conservation of volume facilitates the exchange of information among the branches \cite{kim2015efficient}. The illumination pattern from the optical feedback control is updated in short time intervals, based on the change in shape of \textit{Physarum} as recorded by an overhead camera, which feeds back into updating the neural network. As a result, \textit{Physarum} changes its shape dynamically to maximize nutrient absorption while avoiding broadband light stimuli. 

Iwayama \textit{et al.} analyzed the number of boundaries between distinct spatial regions of increasing and decreasing thickness in the \textit{Physarum} body—termed ``traveling waves'' in their paper—while it was searching for a TSP solution \cite{iwayama2016decision}. They found that fewer ``traveling waves" were hallmarks of the shorter tour lengths.  This suggests that the organism's oscillations are globally coordinated, enhancing its solution-searching ability not through random fluctuations, but through highly synchronized oscillatory dynamics. However, the mechanism by which such highly synchronous oscillations contribute to search performance remains an open question.

The TSP is one of the most extensively studied combinatorial problems.  For a $N-$city problem, the number of distinct feasible tours is $(N-1)!/2$. It is a non-deterministic polynomial time (NP)-hard problem, and no algorithm is known that can derive the exact solution (i.e., the shortest tour) in computational time that scales polynomially with the number of cities. However, many approximate algorithms can quickly obtain high-quality solutions, typically within a few percent of the exact solution. Examples include the Lin-Kernighan algorithm \cite{lin1973effective}, neural network algorithms \cite{hopfield1985neural}, simulated annealing \cite{kirkpatrick1983optimization}, ant colony optimization \cite{dorigo1996ant}, and the Christofides algorithm \cite{Christofides1976WorstCaseAO}.

\textcolor{black}{In computational complexity theory, NP (non-deterministic polynomial time) and P (polynomial time) are two important classes of decision problems. The problems in class P can be solved in polynomial time, whereas a problem belongs to class NP if its solution can be verified in polynomial time. The P vs. NP problem---whether all problems whose solutions can be verified in polynomial time can also be solved in polynomial time--- remains one of the most important open problems in computer science. \textcolor{black}{While \textit{Physarum} finds high-quality solutions to the TSP, we have yet to observe the organism achieving the exact solution for problem sizes $N\geq6$.} Therefore, we do not claim to solve NP-hard problems with \textit{Physarum} but instead seek to provide insights into computational complexity theory and the potential role of living active matter in addressing the $\mathrm{P} \stackrel{?}{=}\mathrm{NP}$ problem.}

In simulated annealing, the goal is to find the global minimum of an energy landscape, which generally corresponds to the ground state of a system. Stochastic resetting \cite{evans2020stochastic,evans2011diffusion} is a specific technique that has been applied to improve the efficiency of such an optimization process. Stochastic resetting can overcome local energy minima that trap the optimization algorithm in unproductive searches, thereby preventing it from finding the global minimum, via “jumps” (resets/collapses) outside of the local neighborhood of exploratory phase space. 
Similarly, in our setup, the optical feedback mechanism -- controlled by a classical, modified Hopfield neural network -- defines an energy landscape as a function of the organism's shape (i.e., the lengths of its branches). \textit{Physarum} acts as an agent that descends the gradient of this landscape while demonstrating the capacity to escape local minima to explore more favorable high-quality solutions.
In other words, the optical feedback mechanism acts as a resetting mechanism that perturbs \textit{Physarum}'s current state, driving the dynamics to escape suboptimal configurations and toward solving a particular TSP instance. 

\textcolor{black}{Although this work reanalyzes the imaging data reported in Ref. \cite{zhu2018remarkable}, the scientific questions addressed and the analyses performed are fundamentally different from those of the previous study. While Ref. \cite{zhu2018remarkable} demonstrated that \textit{Physarum} can generate high-quality approximate solutions to the TSP, the dynamical mechanism by which the organism selects its final solution remained unexplored. Moreover, the lane-state amplitude-based readout of the solution in Ref. \cite{zhu2018remarkable} could only be performed after the optical-feedback pattern had stabilized in the non-equilibrium steady state (NESS), providing no insight into the distinct stages of the solution-search process or the early predictability of the final solution. In contrast, the present work not only investigates the dynamical mechanisms by which \textit{Physarum} reaches a TSP solution and the distinct stages of its solution-search process, but also demonstrates that the solution can be predicted shortly after the onset of optical feedback, well before the NESS is reached. To explore these oscillatory mechanisms, we investigated the spectral and synchronization dynamics underlying the \textit{Physarum} solution-search process to identify early signatures of solution formation, for reliable and more efficient solution prediction.}

\textcolor{black}{While power spectral density (PSD) is a well-established tool for identifying the frequency composition and spectral amplitudes of oscillatory signals, and the Hilbert transform is widely used to extract their instantaneous amplitudes and phases, their application to characterize the evolving dynamics of \textit{Physarum} within the lanes of a slime-machine TSP solver is unprecedented. Likewise, although the Kuramoto order parameter is widely used to quantify phase synchronization in oscillator systems, its application to the experimentally measured phase dynamics of the solution and non-solution lane groups in \textit{Physarum} networks reveals a powerful synchronization metric that characterizes and predicts the selected solution. Together, these complementary analyses provide a quantitative framework for identifying spectral and synchronization-based markers of solution selection, enabling the computation performed by a living biological substrate to be characterized in a manner that was previously inaccessible.}


\textcolor{black}{In this paper, we take a first step toward understanding the physical basis of slime mold solutions to the TSP by investigating the experimentally observable macroscopic dynamics underlying the solution-search process.} We exhaustively analyze TSP solutions obtained by \textit{Physarum} across 41 trials for problem sizes ranging from four to eight. By leveraging the interaction between \textit{Physarum}'s photoavoidance behavior in its branches and illumination patterns updated by a modified Hopfield neural network, we present \textit{Physarum} with TSP instances as environmental constraints that guide its adaptive solution search. Focusing on the high-quality TSP solutions obtained from \textit{Physarum}, we analyze how its oscillation amplitudes, frequency components, and phases correlate with search performance. We employ Fourier transform and power spectral density analyses to extract frequency components of these oscillations across different lanes of the chip, to highlight the distinctions between solution and non-solution lanes. Additionally, we analyze the solution times and compare the scaling of computational time with problem size against silicon-based algorithms. To further investigate the internal dynamics, we analyze the time series of phase variations in \textit{Physarum}'s lanes and evaluate the relevant synchronization indices over subgroups of lanes to assess their correlations at different time periods in the amoeba's progress toward a TSP solution. Our results reveal three distinct dynamical phases in \textit{Physarum}'s behavior while solving the TSP, culminating in the non-equilibrium steady state. Finally, we propose that this synchronization order parameter will serve as a quantitative metric for evaluating \textit{Physarum}’s computational capacity and as an observable that can identify and predict the selected solution shortly after instantiation of the problem. Driving the relative synchronization between the lane groups can coax the system toward higher-quality solutions.

{
\renewcommand{\thesection}{\textcolor{black}{\arabic{section}}}
\section{Results}
}
{
\renewcommand{\thesubsection}{\textcolor{black}{\arabic{section}.\arabic{subsection}}}
\subsection{\textit{Physarum} dynamics and control in the traveling salesperson problem}
}

\begin{figure*}
    \centering
    \includegraphics[width=1\textwidth]{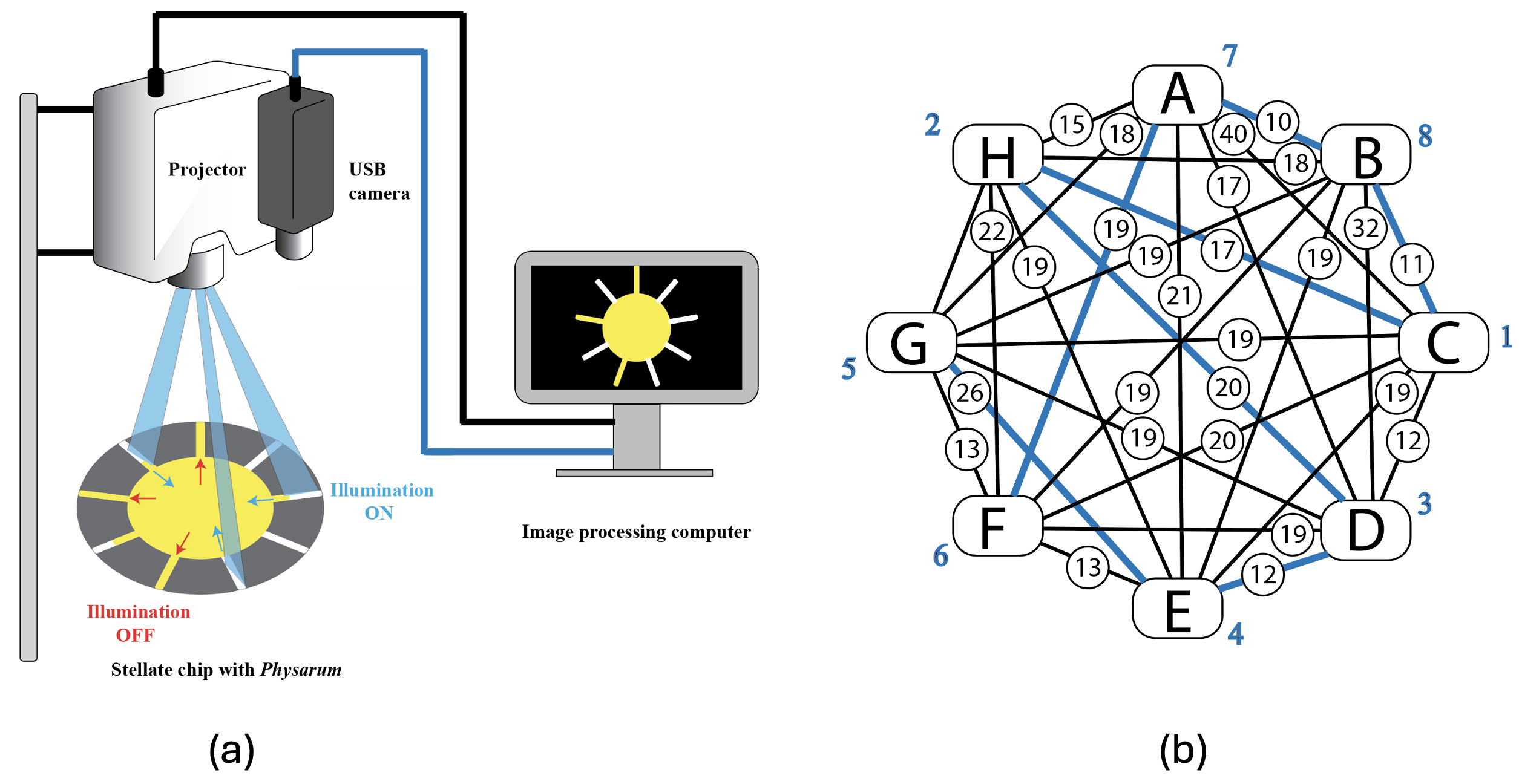}
    \includegraphics[width=1\textwidth]{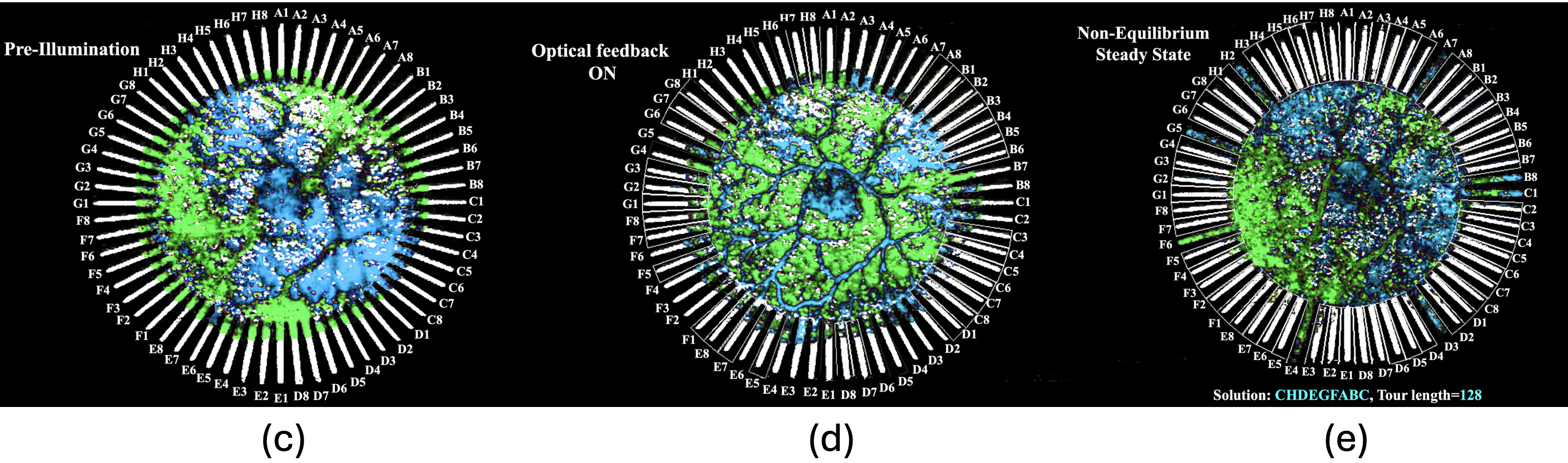}
    \caption{\textbf{\textit{\textit{Physarum} }exhibits microscopic cytosol oscillations and photoavoidance behavior, which enables it to solve the traveling salesperson problem (TSP) under optical feedback control.} (a) Schematic of the experimental setup, illustrating the illumination setup and the balance between inflow and outflow of \textit{Physarum} resources due to its photoavoidance behavior. The optical feedback control from a modified Hopfield neural network stimulates \textit{Physarum} to adapt its oscillatory dynamics, optimizing its growth to solve the TSP.  (b) eight-city TSP instance presented to \textit{Physarum}, which follows a unimodal distribution of tours, as shown in Fig. \ref{Tour_distribution}a.
    Figures 1(c–e) show snapshots of color-coded thickness oscillations in an amoeba resting on a nutrient-rich agar plate and a 64-lane container chip for the eight-city TSP. Bright green and sky blue colors indicate an increase and decrease in out-of-plane thickness, respectively. These oscillations are shown across three phases of \textit{Physarum}’s TSP-solving process: (c) pre-illumination, (d) optical feedback ON, and (e) non-equilibrium steady state (NESS).}
    \label{fig:Scheme&chip}
\end{figure*}

We implemented a technique that utilizes a modified Hopfield neural network to control \textit{Physarum}'s optical feedback loop while solving the TSP \cite{aono2009amoeba,zhu2013amoeba,iwayama2016decision,zhu2018remarkable}. \textit{Physarum} is a complex nonlinear system, with highly synchronized oscillatory dynamics in the presence of considerable thermal noise. Such an approach with stimulus/probe and imaging readout mirrors other implementations for so-called reservoir computing. These include animate systems like brains \cite{cai2023brain} and multicellular collectives \cite{nikolic2023computational}, as well as inanimate systems such as water \cite{fernando2003pattern}, mechanical oscillators \cite{coulombe2017computing}, brain-inspired networks \cite{cucchi2021reservoir}, and photonic systems based on silicon photonic chips, optical delay lines, and photoexcitable devices \cite{van2017advances}. In our setup, the lanes in the chip are labeled as $(V,k)$ where $V \in [1,N]$ represents the city and $k \in [1,N]$ represents the order of visit. Thus, if \textit{Physarum} elongates its branch fully in lane $(V,k)$, it shows that city $V$ is visited in order $k$. As a result, we need stellate chips with $N^2$ lanes to solve a $N$-city problem (Fig. \ref{fig:Scheme&chip}).

{
\renewcommand{\thesubsection}{\textcolor{black}{\arabic{section}.\arabic{subsection}}}
\subsection{Optical feedback control}
}

The state of each lane is denoted by $X_{V k}(t) \in[0,1]$, which is the area occupied by a \textit{Physarum} branch in a lane divided by the total area of the lane. When the branch is fully elongated, $X_{V k} \sim 1$. The illumination for each lane is denoted by $L_{V k}$, whose value is between 0 (no illumination) and 1 (illumination with maximum intensity). The growth of \textit{Physarum} in a branch can be decreased/controlled by setting the illumination $L_{V k}>0.5$. The light stimuli $L_{V k}$ are updated synchronously \cite{zhu2018remarkable}, based on a slightly modified Hopfield-Tank model \cite{hopfield1985neural}:
\begin{equation}
L_{V k}(t+\Delta t)=1-\operatorname{\Theta}\left(\sum_{Ul} W_{V k, U l} \cdot \sigma_{35,0.6}\left(X_{Ul}(t)\right)+0.5\right),
\label{eq:LVk}
\end{equation}
with the application of
\begin{equation}
\sigma_{\alpha, \beta}(x)=1 /\{1+\exp [-\alpha(x-\beta)]\}.
\label{eq:Sigmoid}
\end{equation}
In Eq. \ref{eq:LVk}, $\Theta$ is the Heaviside step function, and $\sigma_{35,0.6}$ is a sigmoidal activation function (as defined in Eq. \ref{eq:Sigmoid}) applied to $X_{V k}$ that tunes sensitivity in the experiment, resulting in enhanced \textit{Physarum} efficiency in solving the TSP. Such an activation function has been confirmed to significantly impact search performance in the algorithmic toy model (see Section 2.2.1), and other activation functions are being explored (see Table S3 of the Supplementary Material).

The symmetric coupling weight matrix $W_{V k, U l}\left(=W_{U l, V k}\right)$ establishes connections between the ``neurons" (in our case, the \textit{Physarum} branches) in lanes $X_{U l}$ and $X_{V k}$. 
\begin{equation}
W_{V k, U l}=\left\{\begin{array}{cc}
-\lambda & \text { if } V=U \text { and } k \neq l \\
-\mu & \text { if } V \neq U \text { and } k=l \\
-\nu \cdot \operatorname{dist}(V, U) & \text { if } V \neq U \text { and }|k-l|=1 \\
0 & \text { otherwise }
\end{array}\right.
\label{eq:Weight_matrix}
\end{equation}
The parameters $\lambda, \mu$, and $\nu$ appearing in the description of the weight matrix $W_{V k, U l}$ in Equation \ref{eq:Weight_matrix} are constants and can vary according to the problem and algorithm. They impose the following constraints on the TSP, respectively: (i) a once-visited city cannot be revisited; (ii) multiple cities cannot be visited simultaneously; and (iii) minimization of the total distance, where $\operatorname{dist}(V, U)$ is the distance between cities $V$ and $U$. For the eight-city TSP experiment, the parameters were chosen as follows: $\lambda=0.5,\mu=0.5$, and $\nu=0.0081$.

The non-Euclidean, non-metric TSP instances for problem sizes ranging from four to eight cities were presented to \textit{Physarum} through modulation of the optical feedback, controlled by the weight matrix of the modified Hopfield neural network. Non-Euclidean instances need not obey Pythagorean distance rules and may or may not be metric (i.e., satisfy the triangle inequality). While non-Euclidean instances—particularly non-metric ones—are generally considered theoretically harder than Euclidean metric problems, this distinction is not strictly black and white. Paradoxically, non-Euclidean TSPs can sometimes be easier for search heuristics such as Lin-Kernighan-Helsgaun (LKH) due to greater variability in pairwise distances. Further details, including comparisons between LKH scaling for Euclidean metric TSP instances and \textit{Physarum} solution scaling for non-Euclidean, non-metric TSP instances, are provided in Section S4 of the Supplementary Material. \textcolor{black}{The TSP map for the non-Euclidean, non-metric, eight-city instance considered here is shown in Fig. \ref{fig:Scheme&chip}b. The tour length is computed by summing the distances between consecutive cities in the tour according to the distance matrix for a corresponding TSP map. Each TSP map was constructed such that the shortest and longest valid tours have total lengths of 100 and 200, respectively. Here, a trial refers to a single independent \textit{Physarum} TSP experiment, each of which yields one \textit{Physarum}-selected tour. A tour denotes any valid route through the cities in the corresponding TSP instance, including tours not selected by \textit{Physarum}.}

Notably, \textit{Physarum} achieved a valid TSP solution, regardless of its quality, 80–90\% of the time across each of the tested problem sizes ($N=4-8$). Moreover, the fraction of high-quality solutions close to the optimal tour length ranged from 20\% in the eight-city TSP ($n_\text{trials}=23$) to 37.5\% in the seven-city TSP ($n_\text{trials}=11$).

The definition of a high-quality tour often depends on the specific solver and practical constraints. Modern heuristic solvers such as the Lin-Kernighan-Helsgaun (LKH) \cite{HELSGAUN2000106} can find near-optimal solutions often within a few percent of the optimal solution. For symmetric TSPs, exact solvers like Concorde \cite{Applegate2006} employ advanced branch-and-cut methods to guarantee optimal solutions, although they have exponential worst-case time complexity. Still, these methods remain remarkably effective for real-world instances with thousands of cities. In contrast, classical heuristics like greedy and nearest-neighbor perform poorly on non-metric problems \cite{johnson1997traveling} — the class of problems we address using \textit{Physarum}, which are more difficult than metric TSPs. These heuristics typically yield solutions that are $2-2.3$ times the optimal length, even for problem sizes that are considered small for such methods. For metric TSPs, the Christofides algorithm provides a guarantee of producing solutions within $1.5$ times the optimal tour length \cite{Christofides1976WorstCaseAO}. However, for arbitrary non-metric TSPs—as discussed earlier in this section—even approximating optimal solutions within any constant factor is NP-hard. \textcolor{black}{Given this challenge, we pragmatically adopt a high-quality solution criterion for the non-metric TSP instances presented to the \textit{Physarum}-based solver, classifying solutions with tour lengths under $1.3$ times the optimum as high-quality.}
This threshold represents a 40\% reduction in the optimality excess criterion ((0.5-0.3)/0.5) compared to the Christofides benchmark. While the Christofides bound applies strictly to metric TSPs, this comparison contextualizes our threshold as a stricter practical benchmark for non-metric problems solved by our \textit{Physarum}-based solver. This definition is appropriate in the context of a biocomputing system like \textit{Physarum}, which -- characteristic of life -- does not search exhaustively for the optimal solution but instead produces feasible, near-optimal solutions. We therefore consider tours within this threshold to be high-quality — not necessarily optimal, but achievable within a reasonable time by an amoeba-based biocomputing system. 

 Additionally, we define a `solution lane' as a lane in the chip that is completely filled by a \textit{Physarum} branch in the non-equilibrium steady state (NESS) region (Fig. 1e) and is identified as part of \textit{Physarum}'s solution to the TSP. In contrast, a `non-solution lane' refers to a lane that is partially filled by a \textit{Physarum} branch before the NESS but becomes completely withdrawn during the NESS. This lane is not chosen as part of the TSP solution. Consequently, there are $N$ solution lanes and $N(N-1)$ non-solution lanes. We will use these terms in the following sections consistently to differentiate between \textit{Physarum}'s selected and non-selected lanes. 

 \textcolor{black}{In one of the trials, \textit{Physarum} solved an eight-city TSP instance with a tour length of $1.17$ times the optimal tour. This value falls within the high-quality solution criterion adopted in this work and indicates competitive performance relative to classical heuristics on non-metric TSP instances.} However, in our analysis, we first illustrate the lane-state and synchronization dynamics using an exemplar solution (tour length 128 for $N=8$), and then statistically validate these observations using aggregated analyses across 41 independent trials with valid TSP solutions spanning problem sizes $N=4-8$. Across these trials, 15 of the 41 ($\sim37\%$) tours are high-quality solutions ($\leq 1.3$ times the optimal), while 30 out of 41 ($\sim73\%$) tours fall within $1.5$ times the optimal, showcasing \textit{Physarum}'s consistent ability to produce quality TSP solutions across trials and distinct problem sizes.

 {
\renewcommand{\thesubsubsection}{\textcolor{black}{\arabic{section}.\arabic{subsection}.\arabic{subsubsection}}}
\subsubsection{Optical feedback control at small intervals}
}
 The efficiency of \textit{Physarum} solution search can be improved by optimizing interactions between the organism and the applied light field. There are at least two ways to accomplish this by directly modifying the optical feedback control: (i) update the illumination pattern at relatively shorter intervals, or (ii) ensure that the illumination state $L_{Vk}(t)$ is a smoothly continuous and differentiable function.
In order to achieve (i), we assume that the illumination pattern is updated in a very short time interval $\Delta t$. 
The equations for the states $X_{Vk}$ of the lanes, according to the continuous Hopfield–Tank model \cite{hopfield1985neural}, are given by
\begin{equation}
\frac{d X_{Vk}}{d t} = -X_{Vk} + \sum_{Ul} W_{Vk, Ul} 
  \ \sigma(X_{Ul})
\label{Eq:dXVk}
\end{equation}
Here, $\sigma(X_{Ul})$ is the sigmoid function. Considering $\Delta t$ to be small, the lane occupancy $X_{Vk}$ at a later time $t + \Delta t$ can be written as:
\begin{equation}
X_{Vk}(t + \Delta t) \approx X_{Vk}(t) + \frac{d X_{Vk}(t)}{d t} \Delta t
\label{Eq:TaylorXVk}
\end{equation}
Substituting Equation ~\ref{Eq:TaylorXVk} into Equation ~\ref{Eq:dXVk}, we obtain

\begin{equation}
X_{Vk}(t + \Delta t) = X_{Vk}(t)(1 - \Delta t) + \left(\sum_{Ul} W_{Vk, Ul} \sigma\left(X_{Ul}\right)\right) \Delta t
\label{Eq:XVk-t+dt}
\end{equation}

The state of the optical feedback control, $L_{Vk}(t + \Delta t) = 1$ if $X_{Vk}(t + \Delta t) < 0.5$ and $L_{Vk}(t + \Delta t) = 0$ if $X_{Vk}(t + \Delta t) > 0.5$. This can be easily formulated using the following function:
\begin{equation}
\eta_{Vk}(t + \Delta t) =
\begin{cases} 
1 & \text{if } X_{Vk}(t + \Delta t) > 0.5 \\
0 & \text{if } X_{Vk}(t + \Delta t) < 0.5 
\end{cases}
\end{equation}
or
\begin{equation}
\eta_{Vk}(t + \Delta t) = {\Theta}(X_{Vk}(t + \Delta t)-0.5).
\end{equation}
Here, $\Theta$ is the Heaviside step function. The expression for $L_{Vk}(t + \Delta t)$ now becomes
\begin{equation}
L_{Vk}(t + \Delta t) = 1 - \eta_{Vk}(t + \Delta t)
\end{equation}
 Alternatively, we can write
\begin{equation}
L_{Vk}(t + \Delta t) = 1 - {\Theta}(X_{Vk}(t + \Delta t)-0.5)
\label{eq:LVk_t+dt}
\end{equation}
Using the value of $X_{Vk}(t + \Delta t)$ from Equation ~\ref{Eq:XVk-t+dt}, we rewrite Equation ~\ref{eq:LVk_t+dt} as
\begin{equation}
\begin{split}
L_{Vk}(t+\Delta t) &= 1 - \operatorname{\Theta} \Bigg( X_{Vk}(t)(1 - \Delta t) + \\
& \quad \left( \sum_{Ul} W_{Vk, Ul} \cdot \sigma\left(X_{Ul}\right) \Delta t -0.5 \right) \Bigg)
\end{split}
\label{Short_update}
\end{equation}
 
Equation \ref{Short_update} defines the rule for updating the optical feedback at small, finite time intervals. With the illumination pattern being updated more frequently, we expect to see an improvement in \textit{Physarum}'s networked response. 

To implement (ii), in principle, $L_{V k}(t)$ can be made a continuous function by slightly modifying the above expression. Instead of using the sharp step function $\Theta$, we can employ activation functions with smoother growth, such as the sigmoid, ReLU, and others (see Table S3 in the Supplementary Material). These alternatives make $L_{Vk}$ continuous within the range [0,1], enabling a more precise encoding of the TSP instance into the illumination pattern. This, in turn, may enhance \textit{Physarum}’s capability to solve the TSP. However, realizing this in an actual experiment is challenging, though not entirely impractical. We would need to employ a dynamic mask (or masks), which could allow the precisely tuned incident light intensity to be controllable at the level of the individual lanes of each TSP stellate chip. 

{
\renewcommand{\thesubsection}{\textcolor{black}{\arabic{section}.\arabic{subsection}}}
\subsection{Evolution of \texorpdfstring{$X_{Vk}$}{XVk} in solution and non-solution lanes}
}

\begin{figure*}
    \centering
    \includegraphics[width=0.8\textwidth]{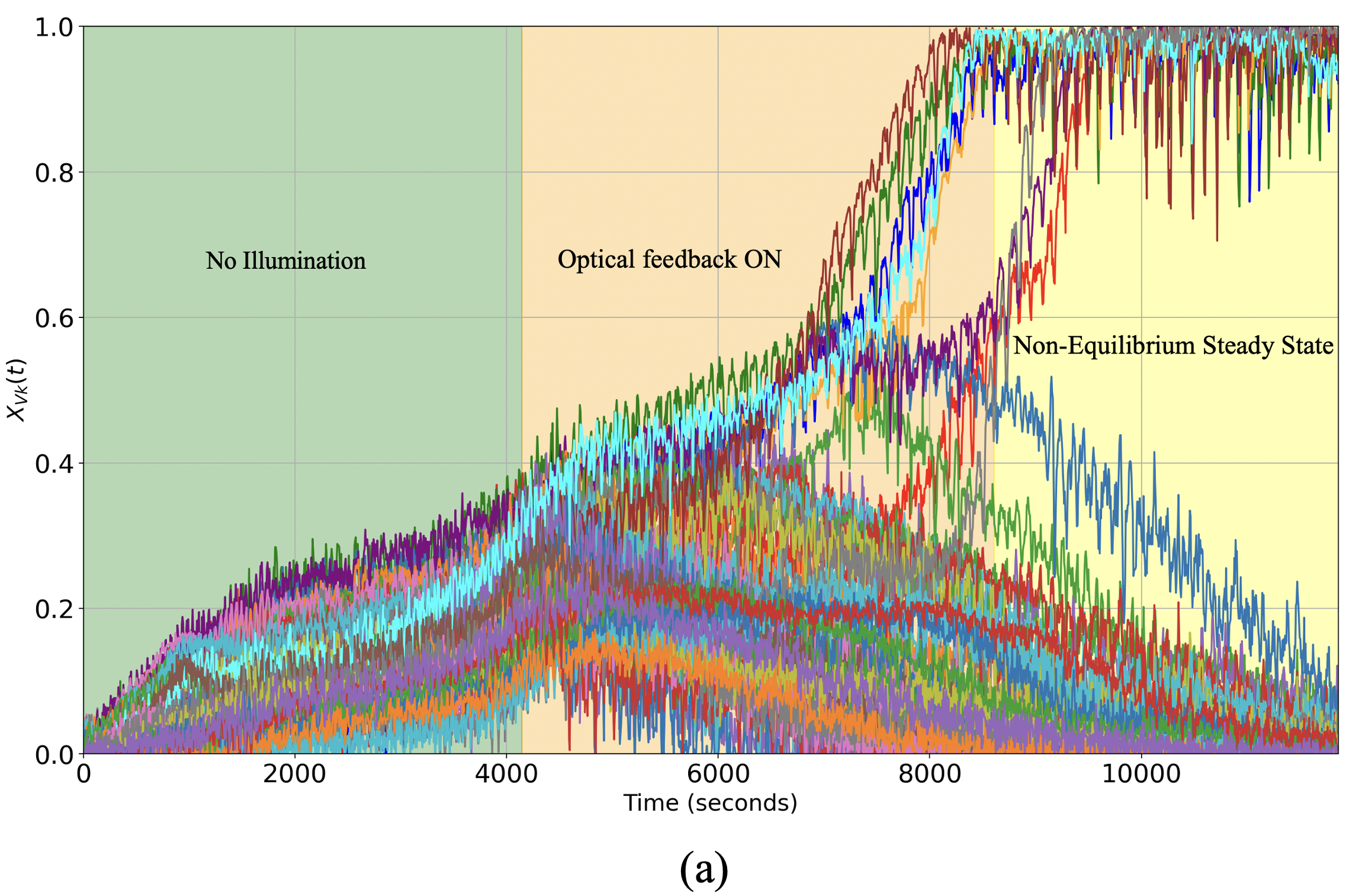}
    \includegraphics[width=0.9\textwidth]{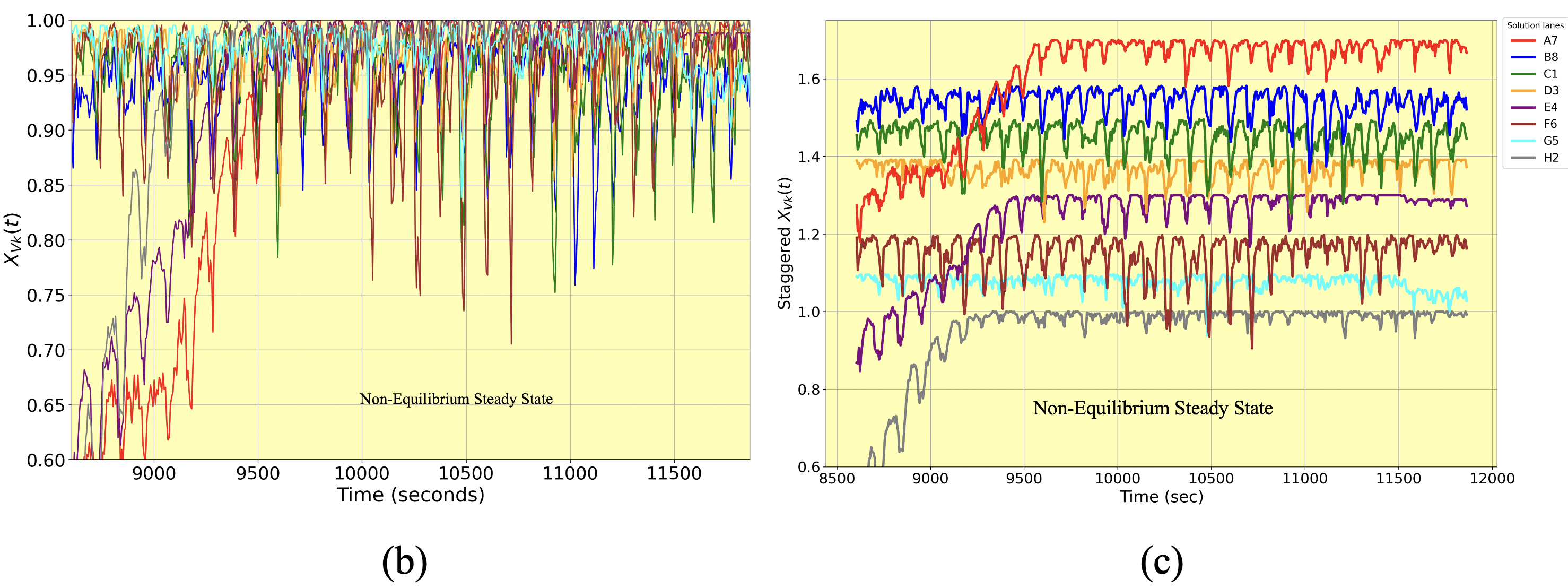}
    \caption{\textcolor{black}{\textbf{Guided by the optical feedback control, \textit{Physarum} finds high-quality solutions to the eight-city TSP, with eight branches fully occupying the solution lanes in the non-equilibrium steady state and exhibiting highly synchronized oscillations, while the remaining branches completely withdraw from the non-solution lanes.} The upper panel shows (a) the evolution of the normalized area occupied by \textit{Physarum} branches ($X_{Vk}$) for all 64 lanes. The lower panels show 
    (b) the multiscale oscillatory $X_{Vk}$ signals of the solution lanes in the non-equilibrium steady state (NESS) from Fig. ~\ref{fig:Scheme&chip}e, which exhibit highly entrained oscillations and motivate the quantitative analysis of their synchronization dynamics, and (c) staggered $X_{Vk}$ values for solution lanes in the NESS from Fig. ~\ref{fig:Scheme&chip}e, showing a higher degree of synchronization than prior time-series segments.}} 
    \label{XVk_All_lanes}
\end{figure*}

Prior to the start of the experiment, \textit{Physarum} is confined to the circular region of the chip, with the lanes completely empty. Once the experiment begins, the illumination pattern ($L_{Vk}(t)$) is updated by referring to $X_{Vk}(t)$ values, according to Eq. ~\ref{eq:LVk}. \textit{Physarum} starts to populate the lanes, causing $X_{Vk}$ values (representing the state of the branches) to increase for all lanes. When $L_{Vk}(t)=1$ (illumination is ON) for a lane, it leads to withdrawal of the \textit{Physarum} branch in the corresponding lane. Consequently, the $X_{Vk}(t)$  values for illuminated lanes ($L_{Vk}(t)=1$) begin to decrease over consecutive updates. Meanwhile, the  $X_{Vk}$  values for unilluminated lanes ($L_{Vk}(t)=0$) continue to increase, eventually approaching 1 near the non-equilibrium steady state (NESS). In the NESS,  $N$  lanes—referred to as solution lanes—achieve  $X_{Vk}$ values close to 1, while the remaining  $N(N-1)$  non-solution lanes have  $X_{Vk}$  values approaching 0. The dynamics observed in both solution and non-solution lanes are shown in Figure \ref{XVk_All_lanes}a.

In the previous study \cite{zhu2018remarkable}, \textit{Physarum} solved the TSP for $4 \leq N \leq 8$. It was shown that \textit{Physarum} successfully found valid solutions in all cases. Additionally, despite the increase in problem size from $N=4-8$, \textit{Physarum} was able to find good-quality solutions in almost linear time. These results suggested that the organism has the ability to search for a reasonably high-quality solution at a low exploration cost, scaling in computation time with a near-linear dependence on the problem size. In contrast, the best approximate algorithms (such as the LKH heuristic, simulated annealing, or genetic algorithm) exhibited only a quadratic dependence at best. 

Noise plays a crucial role in \textit{Physarum}'s adaptation for solving the TSP \cite{folz2021interplay, folz2023noise}. \textit{Physarum}-inspired algorithms fail to converge in its absence; however, simply increasing noise or altering its profile does not necessarily enhance performance and can even degrade solution quality. For further discussion on comparisons of scalings between the experiment, amoeba-inspired TSP algorithms with noise, and other TSP algorithms, see Section S4 of the Supplementary Material.

{
\renewcommand{\thesubsection}{\textcolor{black}{\arabic{section}.\arabic{subsection}}}
\subsection{Synchronization indices}
}
 
 \textcolor{black}{The oscillatory dynamics of \textit{Physarum} branches can be viewed as a network of coupled oscillators. The Kuramoto order parameter \cite{kuramoto1984chemical}, a widely used global measure of phase coherence in coupled-oscillator systems, is therefore a natural choice for quantifying their collective synchronization. Accordingly, we use the synchronization index $S$ to characterize, track, and distinguish synchronization among \textit{Physarum}'s $N$ solution lanes}:


\textcolor{black}{
\begin{equation}
S(t)=\frac{1}{N}\left|\sum_{j=1}^N e^{i \phi_{j}(t)}\right|, 
\label{S}
\end{equation}
where the indices $j$ sum over solution lanes only. In Equation \ref{S}, $\phi_{j}(t)$ is the individual phase of a branch at a particular time, and $S$ is the degree of synchronization among lanes at a specific time.} When there is no synchronization among the phases of lanes, $S$ is 0. Otherwise, if phases of some of the lanes are synchronized, then $S$ takes a positive value, reaching maximum phase synchronization when $S$ is 1.

\textcolor{black}{Taking into account the time dependence of the individual phases $\phi_{j}(t)$ in our time-series data, the time-averaged synchronization index $\overline{S}$, obtained by taking the average of $S$ over discrete timesteps $z=1,2, \ldots, M$ each of duration $\Delta t$, is given by}
\textcolor{black}{
\begin{equation}
\overline{S}=\frac{1}{MN} \sum_{z=1}^M \left|\sum_{j=1}^N e^{i \phi_{j}(z\Delta t)}\right|.
\label{S'}
\end{equation}
$S(t)$ and $\overline{S}$ in Equations \ref{S} and \ref{S'}, respectively, are particularly useful in this context, as we have time-series data on the phases of all lanes, and we want to see to what degree the lanes are synchronized at each timestep, and averaged over a given time window.} 

{
\renewcommand{\thesubsection}{\textcolor{black}{\arabic{section}.\arabic{subsection}}}
\subsection{Determination of the branch phases and synchronization indices of the lanes}
}
\textit{Physarum} is a pulsating organism, and it shows oscillatory dynamics \cite{boussard2021adaptive}. We investigated the phase dynamics of its branches to rationalize its information processing. Every branch of \textit{Physarum} in each chip lane acts as a pseudopod oscillator, with individual oscillators coupled through the central integrated processing body. We computed the instantaneous phases by constructing the corresponding analytic signal. This is done by performing a Hilbert transformation of the original branch state for each lane, $X_{V k}(t)$: 
\begin{equation}
\zeta_{Vk}(t)=X_{V k}(t)+i X_{V k}^H(t)=A_{Vk}(t) e^{i \phi_{Vk}(t)}
\label{eq:Analytic_function}
\end{equation}
$\zeta_{Vk}$ in Eq. \ref{eq:Analytic_function} represents the analytic signal. Here, $A_{Vk}(t)$ and $\phi_{Vk}(\mathrm{t})$ are instantaneous amplitude and phase. $X_{Vk}^H(t)$ is the Hilbert transform of $X_{Vk}(t)$:

\begin{align}
X_{V k}^H(t) &= \frac{1}{\pi} \text{ Principal value } 
\int_{-\infty}^{\infty} \frac{X_{Vk}(\tau)}{t-\tau} \, d\tau \notag \\
&= \frac{1}{\pi} \lim_{\epsilon \rightarrow 0} 
\left(
\int_{-\infty}^{t-\epsilon} \frac{X_{Vk}(\tau)}{t-\tau} \, d\tau 
+ \int_{t+\epsilon}^{\infty} \frac{X_{Vk}(\tau)}{t-\tau} \, d\tau
\right) \label{eq:hilbert_transform}
\end{align}

Now, we can express the time-dependent synchronization indices over solution lanes: 
\textcolor{black}{
\begin{equation}
 S _{\text{Sol}}(t)= \frac{1}{N}\left|\sum_{j=1}^N e^{i \phi_{j}(t)}\right| \label{eq:solution_sync},   
\end{equation}
where the index $j$ sums over solution lanes only. Similarly, using $j$ over the non-solution lanes,
\begin{equation}
S_{\text{Non-sol}}(t)= \frac{1}{N(N-1)}\left|\sum_{j=1}^{N(N-1)} e^{i \phi_{j}(t)}\right| \label{eq:non_solution_sync}.
\end{equation}}

We have computed $S_{\text{Sol}}(t)$ and  $S_{\text{Non-sol}}(t)$ for the entire time-series using Equations \ref{eq:solution_sync} and \ref{eq:non_solution_sync}, respectively, and analyzed their evolution as \textit{Physarum} solves the TSP. The corresponding results are presented in Section 2.8 and Section S5 of the Supplementary Material for various tour lengths and problem sizes.

\subsection{Entrainment in $X_{Vk}$ and Fourier analysis}

As seen in Fig. \ref{XVk_All_lanes}, the evolution of $X_{Vk}$  features two distinct types of dynamics: (1) a sigmoidal-like growth of solution lanes with bifurcation from non-solution lanes, and (2) microscopic cytosol oscillations and other oscillatory behaviors across all lanes. In the NESS region, the  $X_{Vk}$ values for the solution group entrain closely, with the peaks observed in different solution lanes exhibiting strong alignment (Fig. \ref{XVk_All_lanes}, panels b and c). This alignment indicates a higher degree of synchronization among the phases of the solution branches in the NESS. In contrast, the oscillations observed in the non-solution lanes are notably different. The $X_{Vk}$  plots for the non-solution lanes do not entrain, and their oscillations appear out of phase, reflecting a lower degree of synchronization. 

\textcolor{black}{The difference in the synchronization of the $X_{V k}$ signals between solution and non-solution lanes arises from differences in their frequency content. To characterize these oscillatory dynamics, we performed Fourier analyses of the $X_{V k}$ signals, the detailed results of which are presented in the Supplementary Material (Section S2). To facilitate visualization of the low-frequency oscillations, a representative $X_{V k}$ signal was smoothed using a Savitzky-Golay filter (Fig.~S1), while all quantitative analyses were performed on the original, unsmoothed signals. Briefly, the solution lanes exhibit a dominant low-frequency component near ${\sim} 0.01 \mathrm{~Hz}$ with a substantially larger spectral amplitude than the non-solution lanes, whereas the non-solution lanes typically exhibit either a weaker low-frequency peak and/or a dominant higher-frequency component (Fig. S2). These observations motivate the quantitative power spectral density (PSD) analysis of solution and non-solution lane groups presented below.}

\subsection{Power density analysis}
\begin{figure*}
    \centering
    \includegraphics[width=0.55\textwidth]{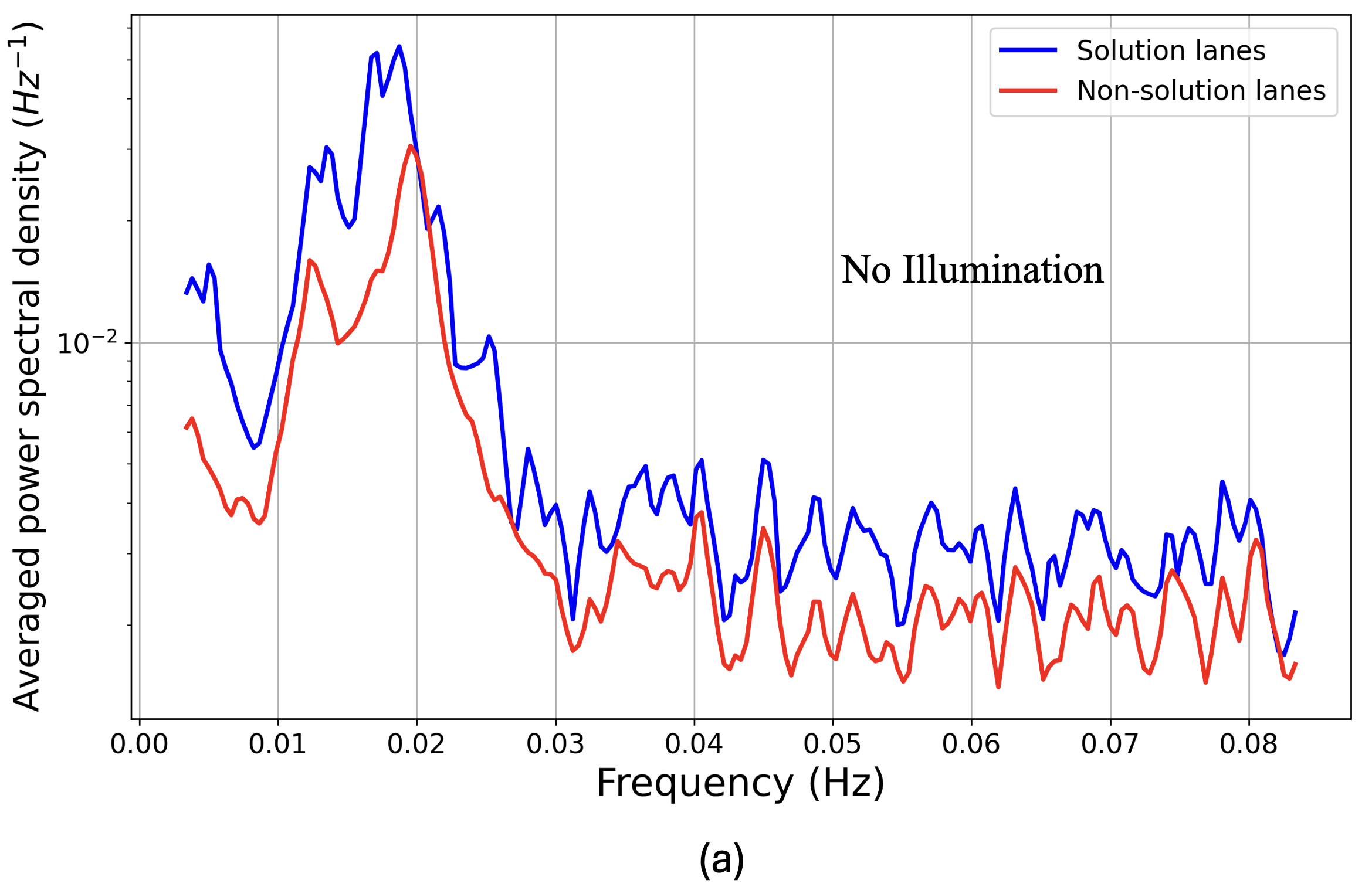}
    \includegraphics[width=0.55\textwidth]{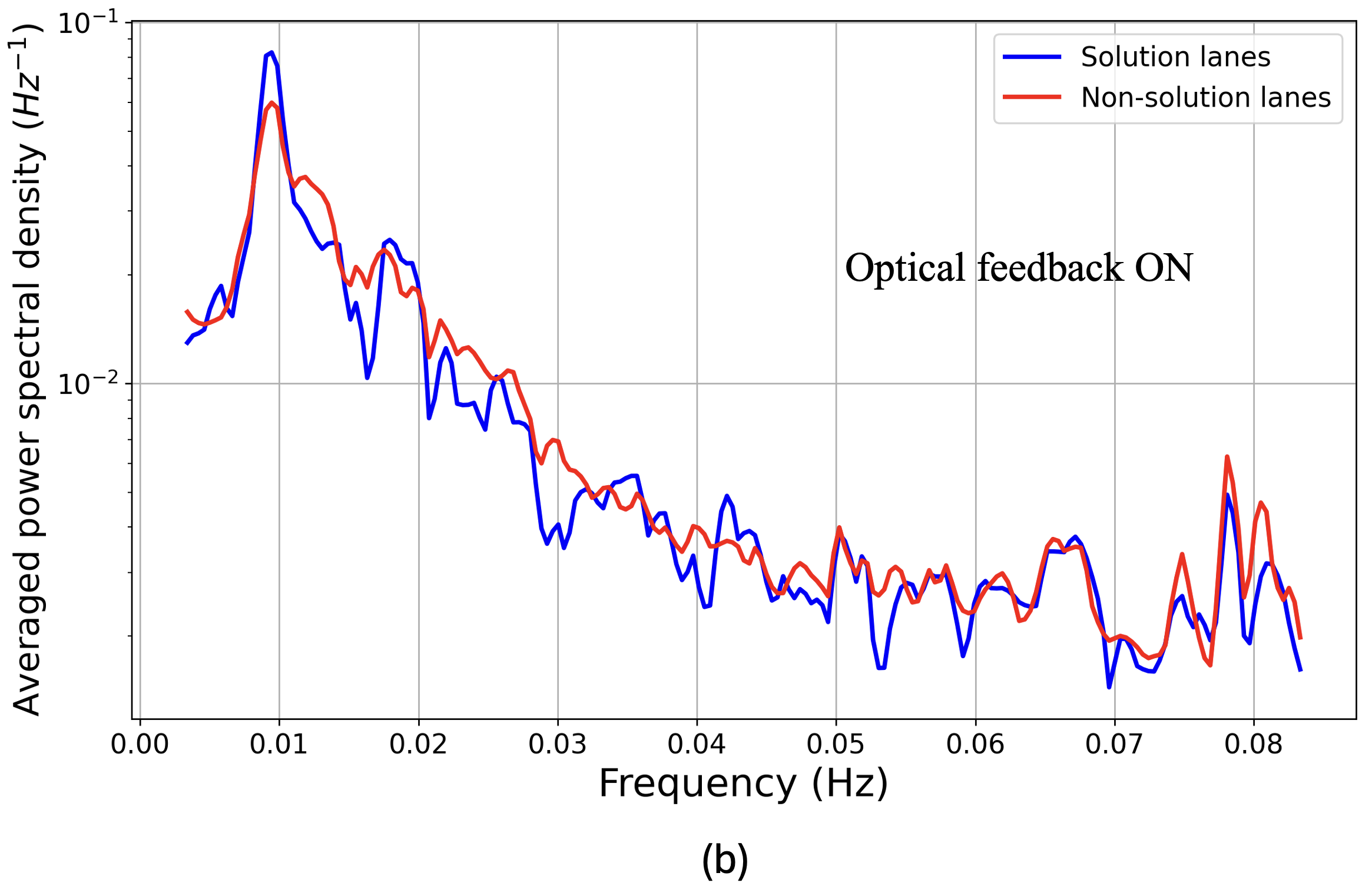}
    \includegraphics[width=0.55\textwidth]{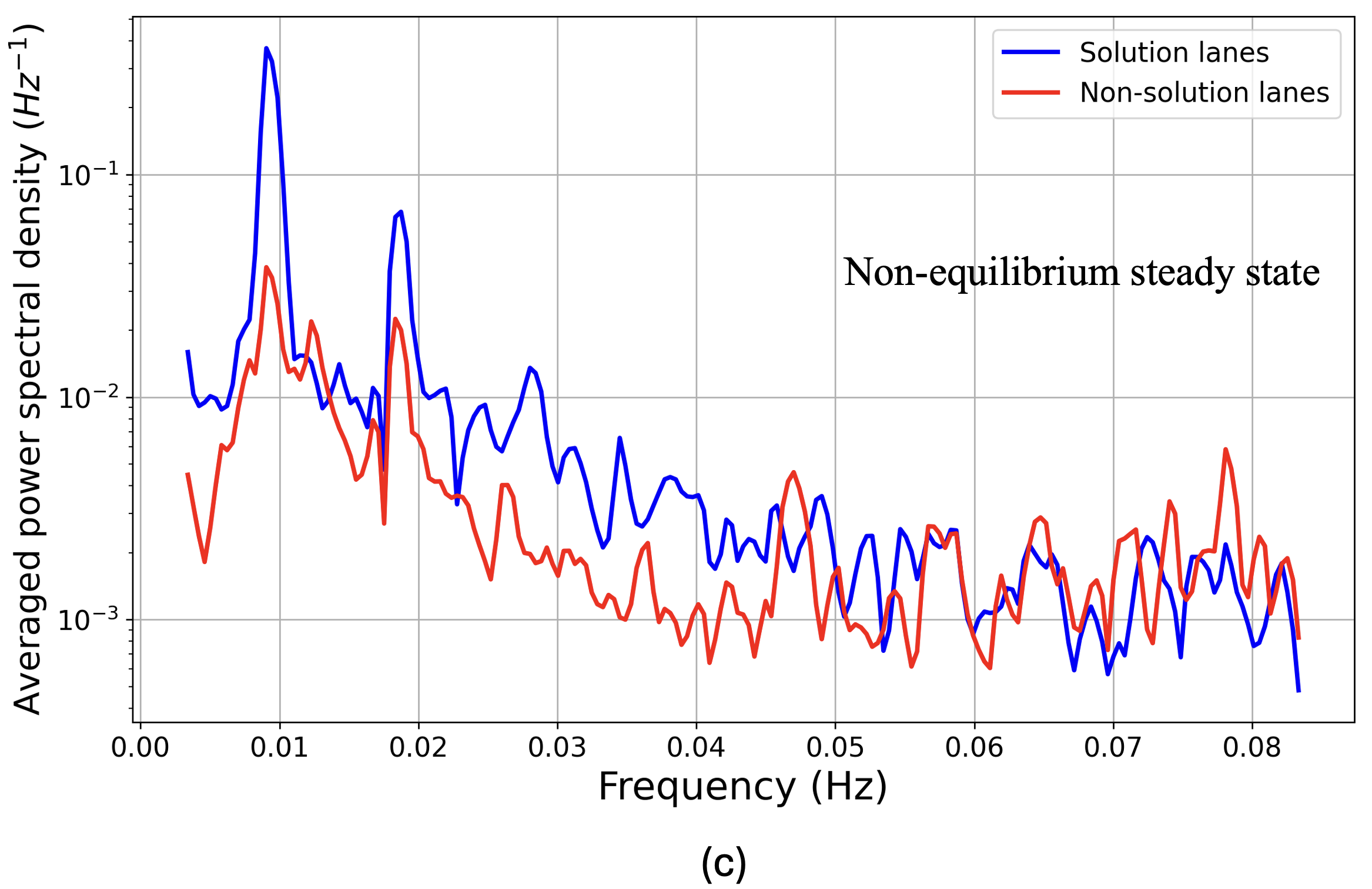}
    \caption{\textbf{The power spectral density (PSD) significantly shifts from the pre-illumination region through the optical feedback to the NESS region, exhibiting marked spectral differences between solution and non-solution lanes in the NESS.} The figure shows the averaged PSD of solution and non-solution lanes in the (a) Pre-illumination region, (b) Optical feedback ON region, and (c) NESS region.}
    \label{PSD}
\end{figure*}

\begin{figure}
    \centering
    \includegraphics[width=1\textwidth]{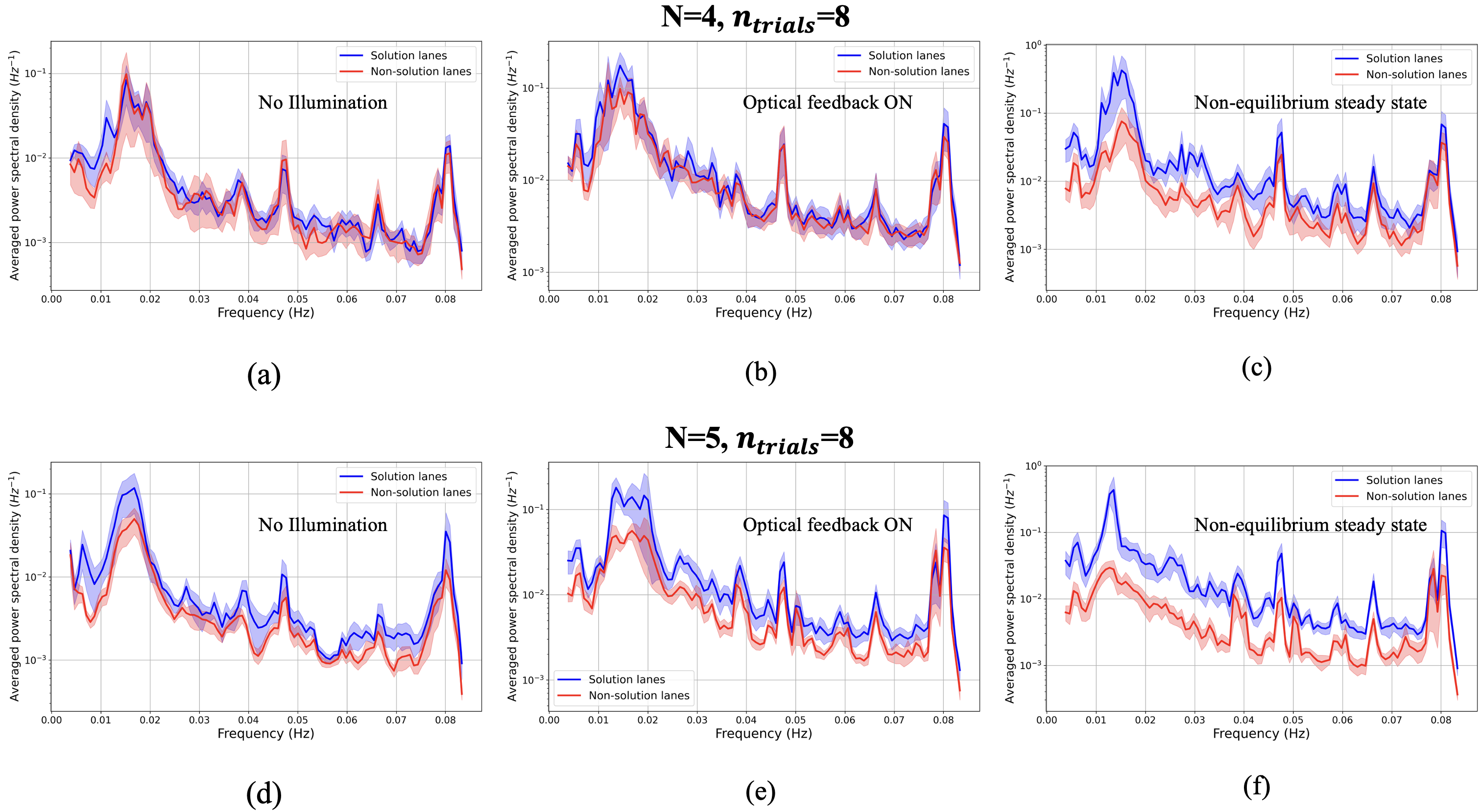}
    \includegraphics[width=1\textwidth]{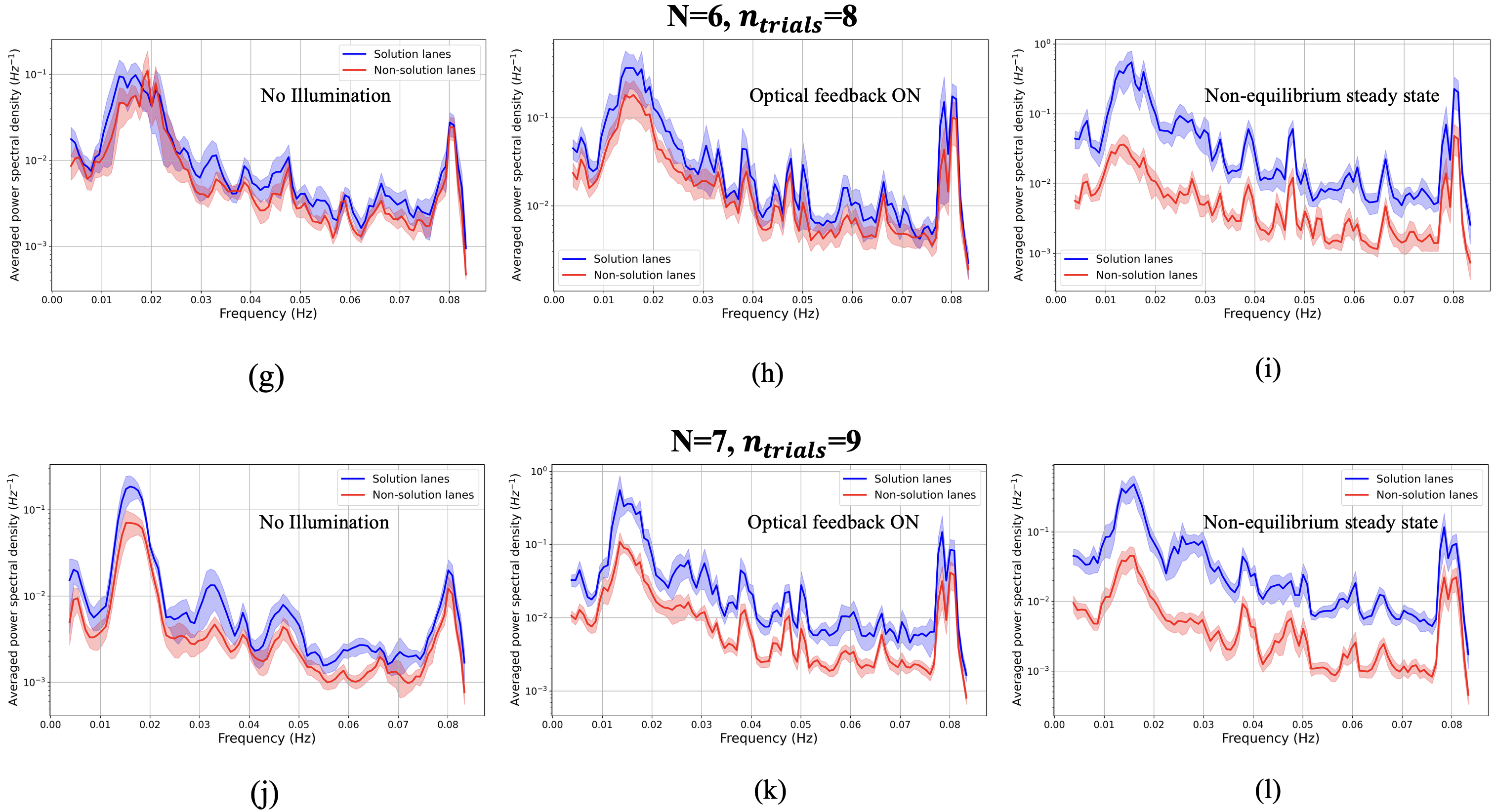}
    \includegraphics[width=1\textwidth]{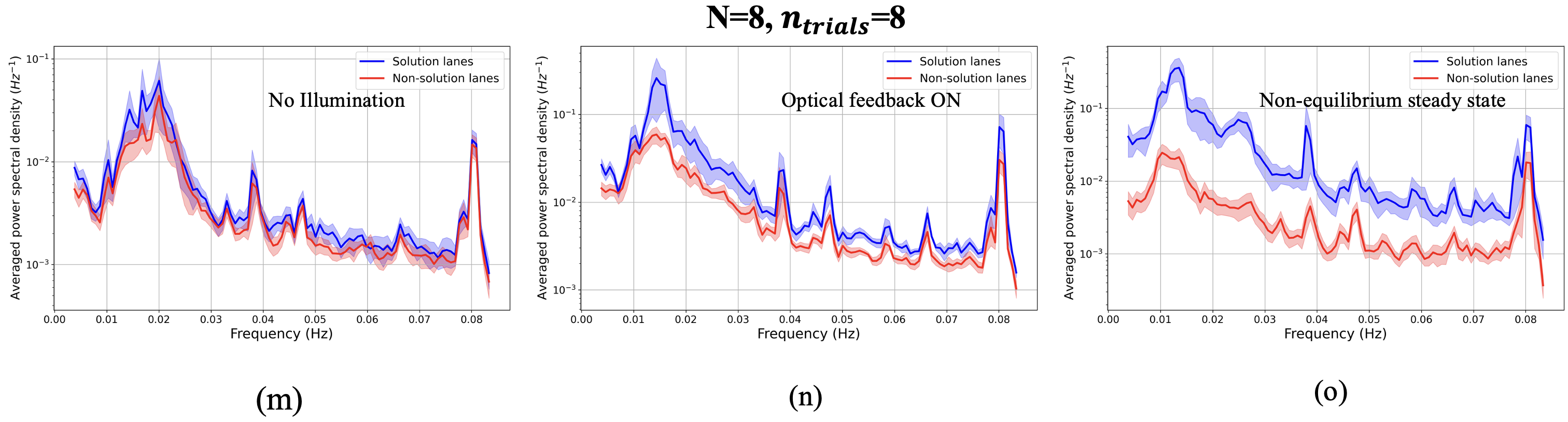}
    \nonumber
\end{figure}
\begin{figure}
    \centering
    \caption{\textbf{Problem size-aggregated PSD signals show a shift in the dominant frequency of the solution lanes from the pre-illumination to the NESS region, along with an amplification gain of the solution-lane PSD signal relative to non-solution lanes, both of which become more pronounced with increasing problem size.} The figure shows problem size-aggregated PSDs for $N=4-8$ across the pre-illumination (left), optical feedback (middle), and NESS (right) regions, showing solution lanes (blue) and non-solution lanes (red). Shaded regions denote $\pm 1$ standard error across trials for each problem size. $n_{\text{trials}}$ denotes the number of independent trials for each problem size.}
    \label{Fig-PSD_Agg}
    \nonumber
\end{figure}
This distinction is further confirmed by the average
power spectral density for solution and non-solution lanes in the NESS region. We used Welch’s method \cite{welch2003use} to compute the power spectral density (PSD) for all lanes. The PSD was then averaged separately for solution and non-solution lanes to obtain their respective averaged power spectral densities. In the pre-illumination (Fig. \ref{PSD}a) and optical feedback regions (Fig. \ref{PSD}b), the PSD signals of solution and non-solution lanes appear similar, with comparable strengths of the most prominent frequency components. However, in the NESS region (Fig. \ref{PSD}c), the strength of the most prominent frequency in solution lanes is amplified compared to the pre-NESS regions, whereas it is suppressed in non-solution lanes. This leads to a clear contrast in the strengths of the most prominent frequency ($\sim$0.01 Hz) components between the two groups in the NESS. Interestingly, the most prominent frequency for both groups shifts across regions: it is $\sim$0.018-0.02 Hz in the pre-illumination region but decreases to $\sim$0.01 Hz in the optical feedback and NESS regions. Similar trends in frequency shifts and signal amplifications are also observed across other tour lengths for $N=8$, as shown in Figure S3 of the Supplementary Material. 

Extending our analysis across multiple trials and problem sizes, we find that these spectral trends are consistently present, albeit with varying magnitude across problem sizes. To demonstrate the statistical robustness of these effects, we plot the problem size-aggregated PSDs for $N=4-8$ in the pre-illumination, optical feedback, and NESS regions (see Fig. \ref{Fig-PSD_Agg}). The shaded regions represent $\pm 1$ standard error across trials for each problem size. $n_{\text{trials}}$ denotes the number of independent trials for each problem size, with 41 trials in total across all problem sizes. We observe that both the frequency shifts and the amplifications of the solution-lane PSD signals are smaller for lower problem sizes and generally become more pronounced as the problem size increases.

To quantify this frequency shift and signal amplification more systematically, we computed the mean normalized frequency shift, $\overline{\Delta \mu}$, defined as the problem size-aggregated mean of the per-trial normalized frequency shifts, $\Delta \mu=\left(\mu_{\text {pre-illumination }}-\mu_{\text {NESS }}\right) / \frac{1}{2}\left(\mu_{\text {pre-illumination }}+\mu_{\text {NESS }}\right)$. Here, $\mu_{\text {pre-illumination }}$ and $\mu_{\text {NESS }}$ denote the dominant frequencies of the solution lanes in the low-frequency part (5-40 mHz) of the spectrum for the pre-illumination and NESS intervals, respectively. The resulting bar plot of $\overline{\Delta \mu}$ as a function of problem size is shown in Fig. S4a of the Supplementary Material, where the values are aggregated over all trials for each problem size and the error bars represent $\pm 1$ standard error. Similarly, we computed the problem size-aggregated mean signal amplification gain, denoted by ${\overline{G}_{\text {NESS}}}$, of solution lanes relative to the non-solution lanes in the NESS region for the low-frequency portion of the spectrum. The corresponding dependence of ${\overline{G}_{\text {NESS}}}$ on problem size is shown in Fig. S4b of the Supplementary Material.

We observe that both the normalized frequency shift $\overline{\Delta \mu}$ and the signal amplification gain $\overline{G}_{\text {NESS }}$ generally increase with the problem size. In the current setup, the total number of lanes is $N^2$, of which only $N$ correspond to solution lanes, while the remaining $N(N-1)$ are non-solution lanes. As the problem size increases, the number of non-solution lanes grows rapidly, leading to a larger fraction of lanes being illuminated under optical feedback, particularly in the NESS region. These illuminated non-solution lanes act as sources of optical drive, injecting energy into the system, which is subsequently redistributed through \textit{Physarum}'s CPU (the central hub of the chip) to the solution lanes. This pool of non-solution lanes grows with increasing problem size, enabling greater energy redistribution toward the solution lanes. The redistributed energy contributes to the amplification of the solution-lane PSD signal relative to that of the non-solution lanes. This amplification increases with problem size, as the pool of non-solution lanes grows. This is consistent with the observed mean amplification gain for the low-frequency region, $\overline{G}_{\text {NESS}}$ (Fig. S4b in the Supplementary Material), which increases from about $ 10$x for $N=4$ to about $43$x for $N=8$, giving a ratio of $\sim 4.3$. Over the same range, the total number of non-solution lanes increases from 12 to 56, corresponding to a ratio of $\sim 4.7$, which is in good agreement with the observed amplification scaling.

This low-frequency amplification of the solution-lane PSD in the NESS is accompanied by a downward shift of the dominant frequency relative to the pre-illumination region, consistent with redistribution of spectral weight under energy conservation. 
\textcolor{black}{Although the frequency shift exhibits non-monotonic behavior at very small $N$, it scales as $N-1$ for moderate $N$ before saturating at large $N$, following the scaling of the amplification gain per solution lane.} This is in agreement with the observed mean frequency shifts at larger problem sizes: for $N=8$, the mean normalized shift is about $0.27$, whereas for $N=6$ it is about $0.18$, giving a ratio of $\sim1.5$, which closely matches the ratio of the corresponding ($N-1$) values, $7/ 5=1.4$. The  mean normalized frequency shifts $\overline{\Delta \mu}$ and the mean solution-lane signal amplification gains $\overline{G}_{\text{NESS}}$ in the low-frequency region are provided in Table S1 of the Supplementary Material.

We provide a simple physical interpretation of this behavior by considering a heuristic toy model in which each lane is treated as an effective oscillator. For simplicity, we assume that the spectral energy of each oscillating branch has contributions only from two dominant modes: a low-frequency mode ($\omega_{\text{low}} \sim0.01-0.02 \mathrm{~Hz}$) and a higher-frequency mode ($\omega_{\text {high}}\sim 0.08 \mathrm{~Hz}$). The effective energy of a branch is then taken to be proportional to $\alpha_{\text {low }} \omega_{\text {low }}^2+\alpha_{\text {high }} \omega_{\text {high }}^2$, where the normalization by the moment of inertia is assumed to be the same for all branches. $\alpha_{\text {low }}$ and $\alpha_{\text {high }}$ are the strength of the PSD signals at $\omega_{\text {low }}$ and $\omega_{\text {high }}$, respectively. Under these assumptions, we estimate the total energy of all branches in the pre-illumination and NESS regions. 

\textcolor{black}{During the optical-feedback interval, most non-solution lanes become optically driven, resulting in an effective energy input of order $N(N-1) \bar{\epsilon}$, where $\bar{\epsilon}$ is the mean energy injected per non-solution lane over the entire interval. All energies in the ensuing discussion have been normalized by angular momentum, which is assumed constant for each branch in our toy model. Consequently, the redistributed optical energy available to each solution lane in the NESS is $e_{\text{in}}=(N-1) \bar{\epsilon}$ on average. The absorbed energy available per solution lane is thus given by
\begin{equation}
e_{\mathrm{abs}}=\min \left(e_{\mathrm{in}}, e_{\mathrm{cap}}\right)=e_{\text{cap}} \min \left(\frac{(N-1) \bar{\epsilon}}{e_{\mathrm{cap}}}, 1\right),
\end{equation}
such that the absorbed energy saturates at $e_{\text{cap}}$.
Imposing the energy-balance condition for all $N$ solution lanes, $E_{\text {pre}}+Ne_{\text{abs}}=E_{\text{NESS}}$, we obtain a scaling relationship between the normalized frequency shift $\Delta \mu_\text{toy}$ of the low-frequency mode and the problem size $N$ as
\begin{equation}
\begin{split}
\Delta \mu_{\text{toy}}=\frac{2}{\omega_{+}^2}\left[\left\{\frac{G_{\text {NESS }}}{k N}+\frac{1}{k}\left(1-\frac{1}{N}\right)-1\right\} \omega_{\text {NESS,low}}^2+\frac{\omega_{\text {NESS,high }}^2}{k}\right]\\
-\frac{2}{\omega_{+}^2}\left[{\frac{e_{\text{abs}}}{\alpha_{\mathrm{pre,low}}N}}+\tilde{\alpha}_{\text {pre }} \omega_{\text {NESS,high }}^2\right],
\label{eq:Delta_mu}
\end{split}
\end{equation}
with $\omega_{+}=\omega_{\text {pre,low}}+\omega_{\text{NESS,low }}$. ${G}_{\text {NESS}}$ represents the amplification gain of solution lanes relative to the non-solution lanes. We define $\tilde{\alpha}_{\text {pre}}\equiv\frac{\alpha_{\text {pre,high}}}{\alpha_{\text {pre,low}}}$ as the ratio of the peak PSD signals at high and low frequencies in the pre-illumination region. The parameter $k\equiv\frac{\alpha_{\text {pre,low}}}{\alpha_{\text {non-sol}}}$ denotes the ratio of the low-frequency PSD signal peak in the pre-illumination region to that of the non-solution lanes in the NESS. A detailed derivation of $\Delta \mu_{\text{toy}}$ is presented in Section S3.1 of the Supplementary Material. Using representative values of $\alpha, \omega$, and $\bar{\epsilon}$ and assuming that the amplification gain $G_{\mathrm{NESS}}$ scales with the absorbed optical energy available to the solution lanes, we compute the dependence of $\Delta \mu_{\text {toy}}$ on the problem size $N$ using Eq. ~\ref{eq:Delta_mu} (see Fig. S6 in the Supplementary Material).
The shift exhibits a non-monotonic dependence at small problem sizes, followed by an approximately linear increase consistent with the observed experimental values, before gradually approaching saturation with increasing $N$.}

\begin{figure}[h]
    \centering
    \includegraphics[width=0.8\textwidth]{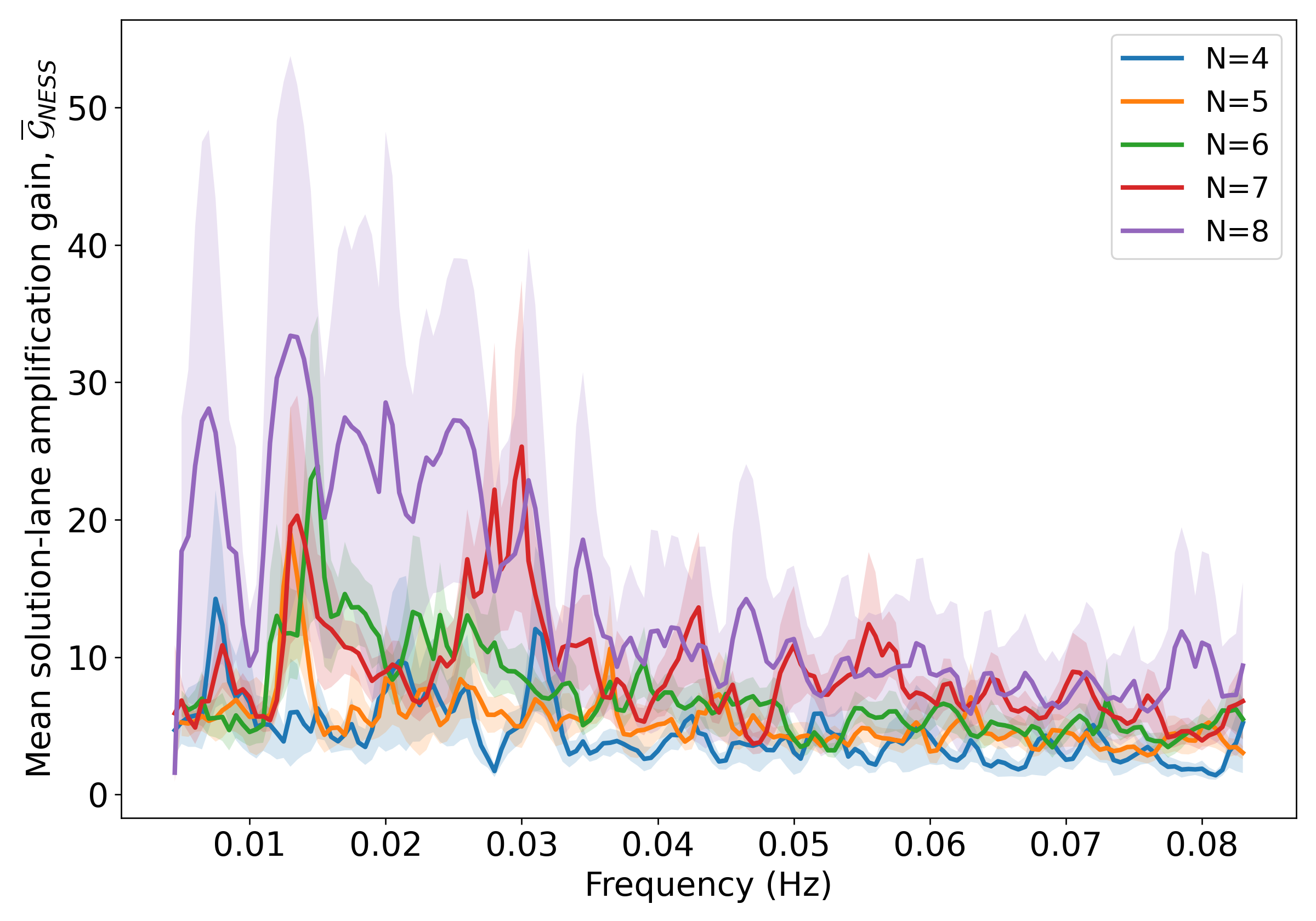}
    \caption{\textbf{The mean frequency-dependent amplification gain of the solution-lane PSD signal in the NESS, $\overline{\mathcal{G}}_{\text {NESS }}(f)$, exhibits pronounced peaks in the low-frequency region $(5-40 \ \mathrm{mHz})$, with the dominant peaks for most problem sizes confined within a narrower range of $8-16$ mHz, consistent with the unperturbed cytosolic oscillatory dynamics of \textit{Physarum}.} Curves represent the $\overline{\mathcal{G}}_{\text {NESS }}(f)$, across trials for each problem size, with shaded regions indicating $\pm 1$ standard error. While the overall low-frequency-dominated spectral structure is preserved across problem sizes, slight shifts in the dominant peak positions are observed, arising from variations in lane configuration, and initial conditions.}
    \label{Fig-Sol_lane_PSD_Amp}
\end{figure}

We next examine how the ratio of the mean PSD of the solution lanes to that of the non-solution lanes varies across all frequencies in the NESS region. The resulting frequency-dependent amplification gain in the NESS region, denoted by $\mathcal{G}_{\text {NESS }}(f)$, is shown in Fig. \ref{Fig-Sol_lane_PSD_Amp} for problem sizes $N=4-8$. Each curve represents the mean amplification gain ($\overline{\mathcal{G}}_{\text {NESS }}(f)$) across all trials for a given problem size, with shaded regions indicating $\pm 1$ standard error. The amplification gain is strongest in the low-frequency regime $(\sim 5-40  \ \mathrm{mHz})$, where most problem sizes exhibit a structured band of multiple amplification gain peaks rather than a single isolated maximum. The dominant peak for most problem sizes occurs within a narrower range of $\sim 8-16 \ \mathrm{mHz}$, consistent with the frequency range of unperturbed, cytosolic oscillatory dynamics of \textit{Physarum} and with additional pronounced peaks at nearby, slightly shifted frequencies in the low-frequency regime. In the $N=7$ configuration, a strong peak near $\sim 10 \ \mathrm{mHz}$ persists, but the dominant peak occurs at a slightly higher frequency ($\sim 30 \ \mathrm{mHz}$), indicating an alteration in the dominant amplification mode while preserving the overall similar low-frequency structure. This alteration may arise from variations in the oscillator dynamics introduced by the experimental implementation of the seven-city problem. In particular, the eight-city chip was used to implement the $N<8$-city experiments by blocking $64-N^2$ lanes, which results in an asymmetric configuration of the active lanes for the odd problem sizes. As this alteration is not observed in the five-city problem, further experiments are necessary to understand how odd problem sizes scale in general compared to even problem sizes.

The observation of such an energy redistribution from collective modes with higher frequency to the low-frequency regime has been studied in diverse protein systems, including bovine serum albumin \cite{nardecchia2018out,azizi2023examining, perez2025unveiling, sushko2015sub}, lysozyme \cite{preto2017semi}, and myoglobin \cite{sushko2015sub}  as well as in other toy models \cite{zhang2019quantum}, evoking the features of a Fr\"{o}hlich condensate \cite{frohlich1968long, wu1978bose, preto2017semi, turcu1997generic}. In the sub-terahertz (THz) range, diverse protein solutions can exhibit enhanced absorption at low concentrations (so-called ``terahertz excess''), a phenomenon attributed to the dynamic activity of proteins in dilute conditions \cite{sushko2015sub}. These solutions also feature extended hydration layers, which modulate protein-water interactions and contribute to low-frequency vibrational dynamics. As modelled atomistically in a recent work \cite{azizi2023examining} from one of the authors (PK) and coworkers, in more concentrated solutions, intimate protein–water interactions can give rise to attractive electrodynamic behaviors among proteins, due to the emergence of a larger electric dipole moment (up to $\sim$1000 debye) across each protein, when decorated by recruited water and ions. Such large, oscillating electric dipoles emit electromagnetic fields like nanoantennas and can synchronize architectonic organization at thermal equilibrium, and considerably more so in the far-from-equilibrium environments of living systems. Most recently, such behavior has been observed in light-harvesting red phycoerythrin protein with either (coherent) laser or (incoherent) thermal excitations \cite{perez2025unveiling} stimulating concentration of energy in low-frequency vibrational breathing modes through non-linear couplings in the protein matrix. While in a frequency regime that is far slower than the THz vibrational modes of individual proteins, the behavior of \textit{Physarum} while solving the TSP under optical feedback control may be suggestive of an organismal-scale Fr\"{o}hlich-like energy redistribution, although direct experimental verification of such a mechanism remains an open question for future work.


\subsection{Analysis of synchronization in branch phases}
\begin{figure}
    \centering
    \includegraphics[width=0.6\textwidth]{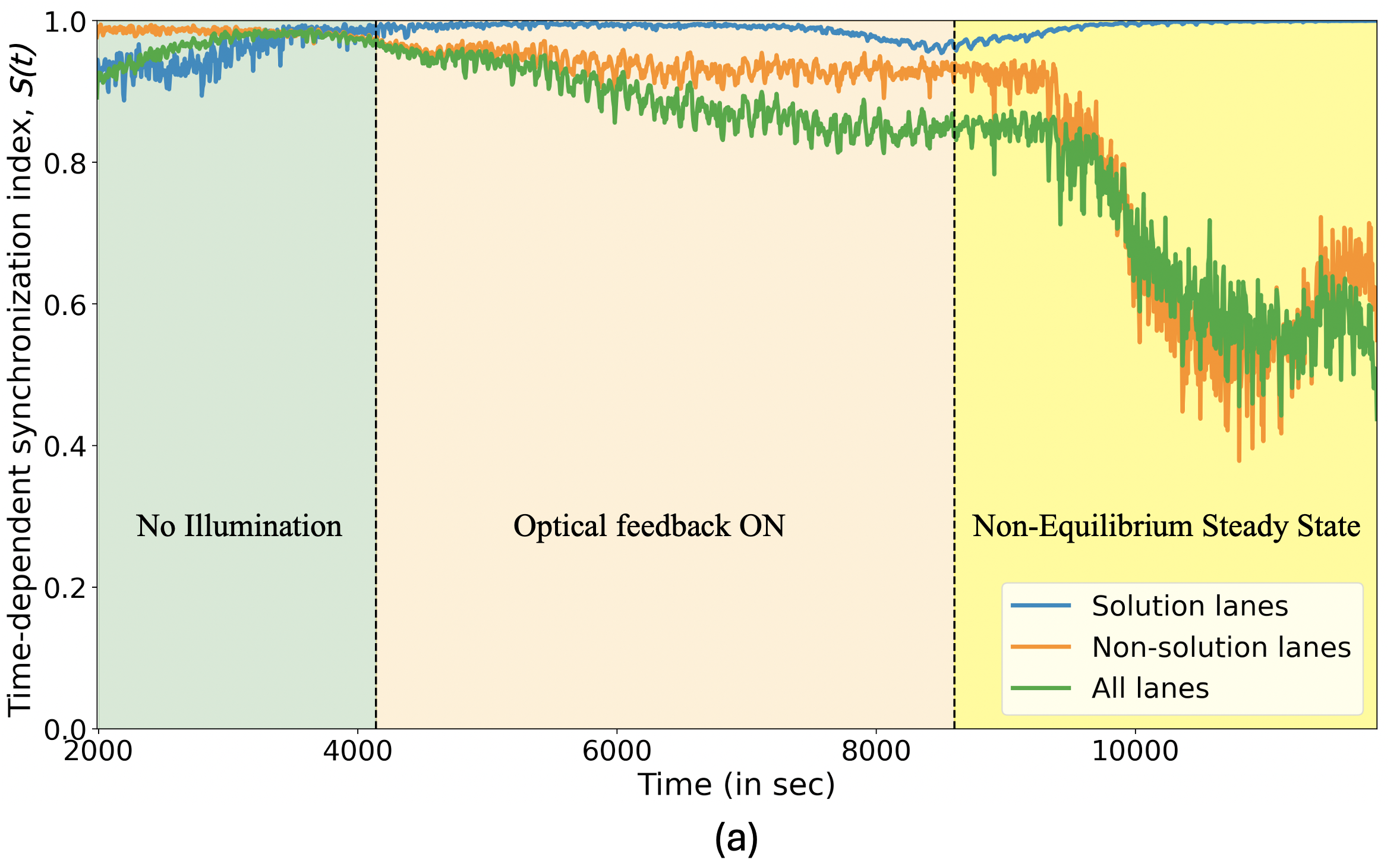}
    \includegraphics[width=0.6\textwidth]{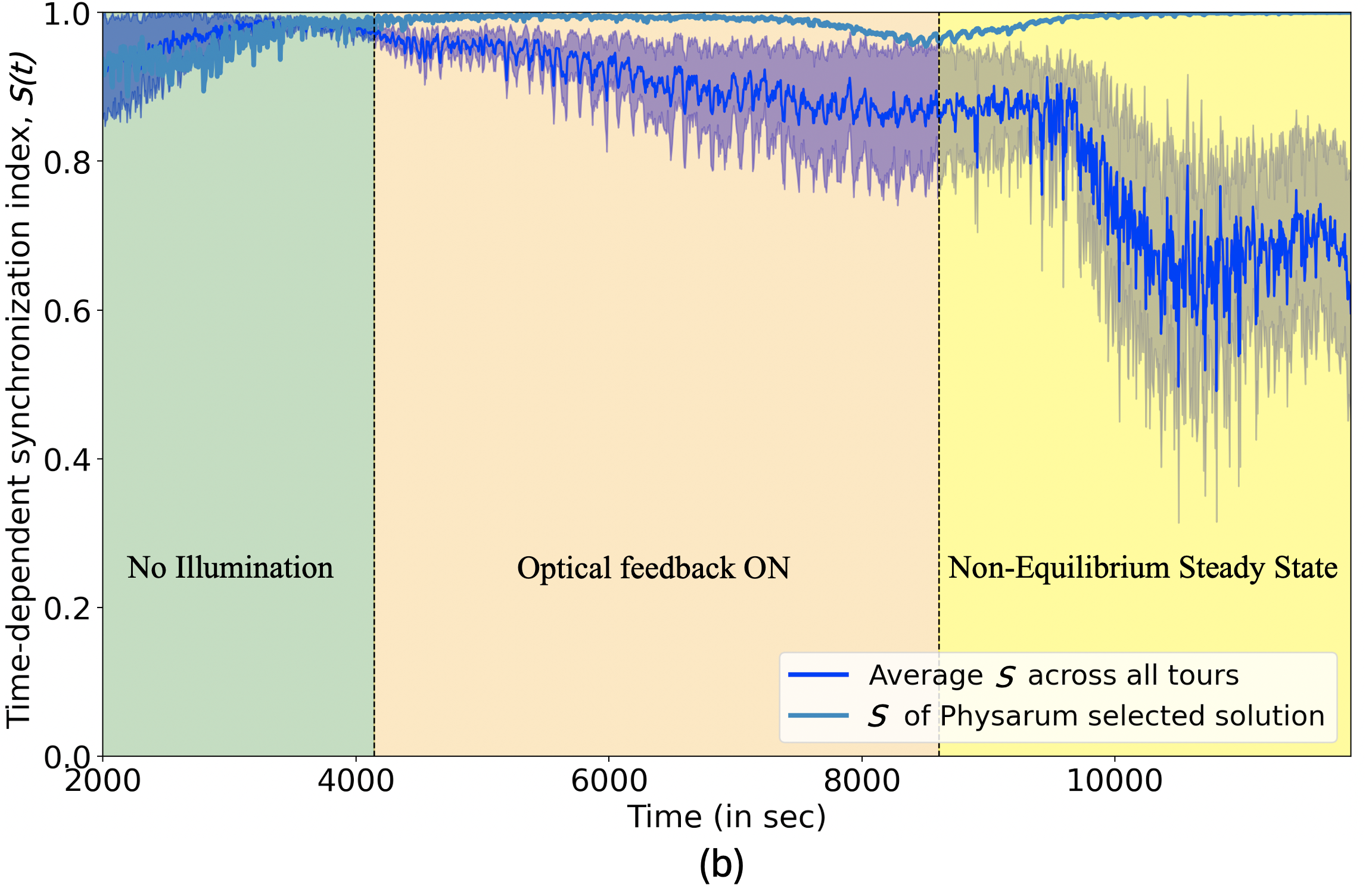}
    \centering
    \caption{\textbf{Optical feedback activation stimulates bifurcation of the synchronization indices ($S(t)$) between solution and non-solution lanes.} In the non-equilibrium steady state (NESS), the $S$ values for the solution lanes reach nearly the maximum ($\sim$1), whereas non-solution lanes exhibit a minimum $S$ value, highlighting the effect of optical feedback in guiding the system dynamics toward the solution. (a) The panel shows time-dependent synchronization indices $S(t)$ given in Equations \ref{eq:solution_sync} and \ref{eq:non_solution_sync} for solution lanes (blue) and non-solution lanes (orange), respectively, and all lanes (green) summed over $Vk$ up to $N^2$ for problem size $N=8$ and tour length of 128. The plot is divided into three distinct regions, as described in the main text. (b) The panel compares the time-dependent $S(t)$ of the selected \textit{Physarum} solution lanes (same data series in blue from panel a) and the average $S(t)$ of non-selected \textit{Physarum} lanes across all 44 distinct possible tour configurations with a tour length of 128 (royal blue). The transparent shaded region around the royal blue data series represents ±1 standard deviation in $S(t)$ values across all these distinct possible tours of tour length 128.
    }
    \label{TLSI}
\end{figure}

To analyze the synchronization dynamics observed in the solution and non-solution lanes, we computed the time-dependent synchronization indices, $S(t)$, and visualized their evolution using time-series plots (Fig. \ref{TLSI}). Based on \textit{Physarum}'s dynamic photoresponsive behavior, the entire time series can be divided into three distinct regions:

\begin{enumerate}
    \item The \textbf{pre-illumination} region, indicated as ``no illumination'' in Fig. \ref{TLSI} (shaded in green), is the period before the optical feedback control is activated. During this time, all lanes exhibit similar  $S(t)$ values, as \textit{Physarum} follows its intrinsic unperturbed dynamics.

    \item The \textbf{optical feedback} region, indicated as ``optical feedback ON'' in Fig. \ref{TLSI} (shaded in beige), begins with the activation of the optical feedback under modified Hopfield network control, during which the TSP instance defined by the Hopfield weight matrix is presented to \textit{Physarum}, encoded in the updating illumination pattern. 

    \item The \textbf{non-equilibrium steady state (NESS)} region (shaded in yellow) is the interval when the illumination pattern stabilizes and \textit{Physarum} reaches a NESS, with its branches in a configuration of fully illuminated ($L_{Vk}(t)=1$), completely retracted ($X_{Vk}(t)=0$) non-solution lanes and unilluminated ($L_{Vk}(t)=0$), completely elongated ($X_{Vk}(t)=1$) solution lanes.
\end{enumerate}

\begin{figure}
    \centering
    \includegraphics[width=0.6\textwidth]{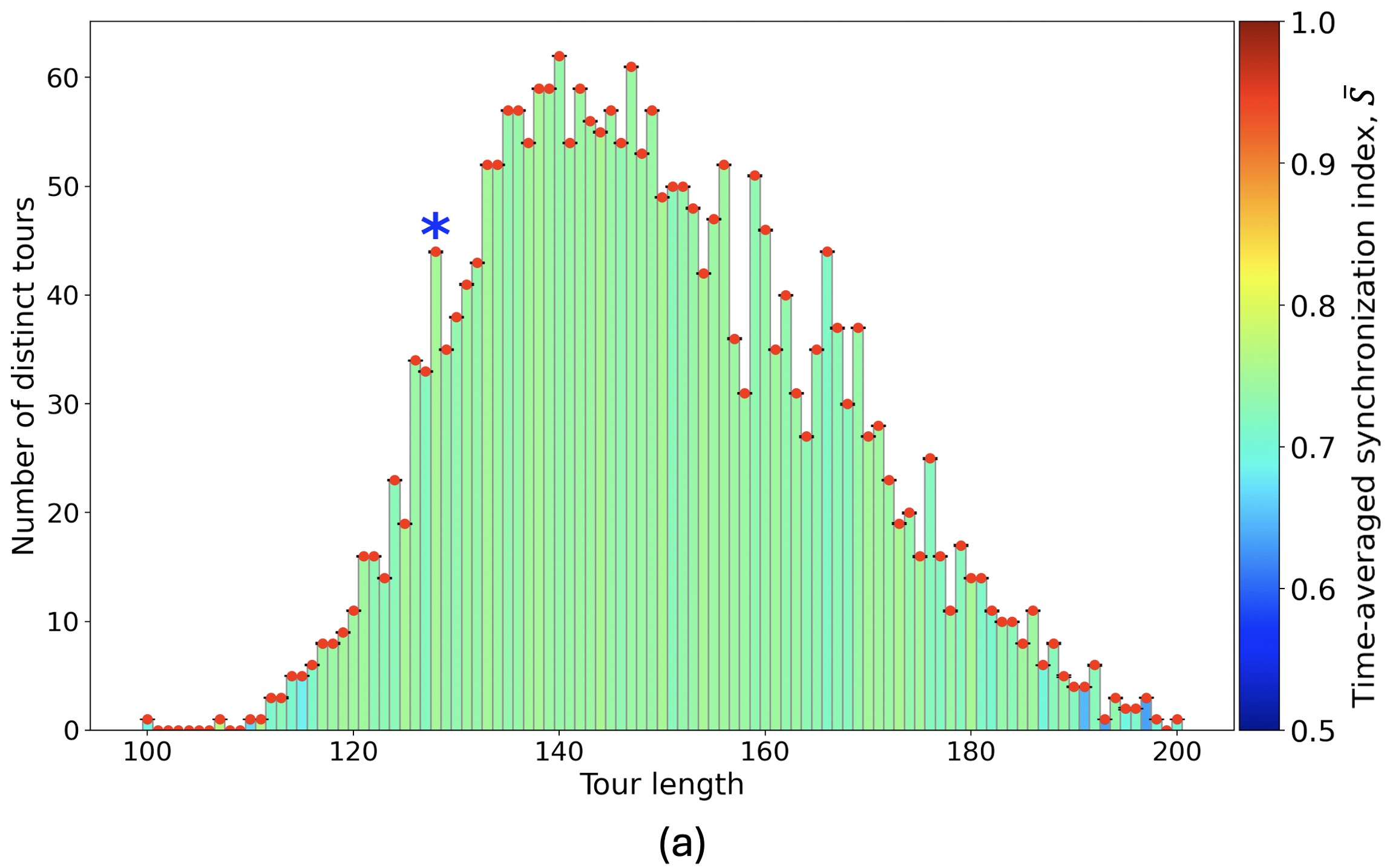}
    \includegraphics[width=0.6\textwidth]{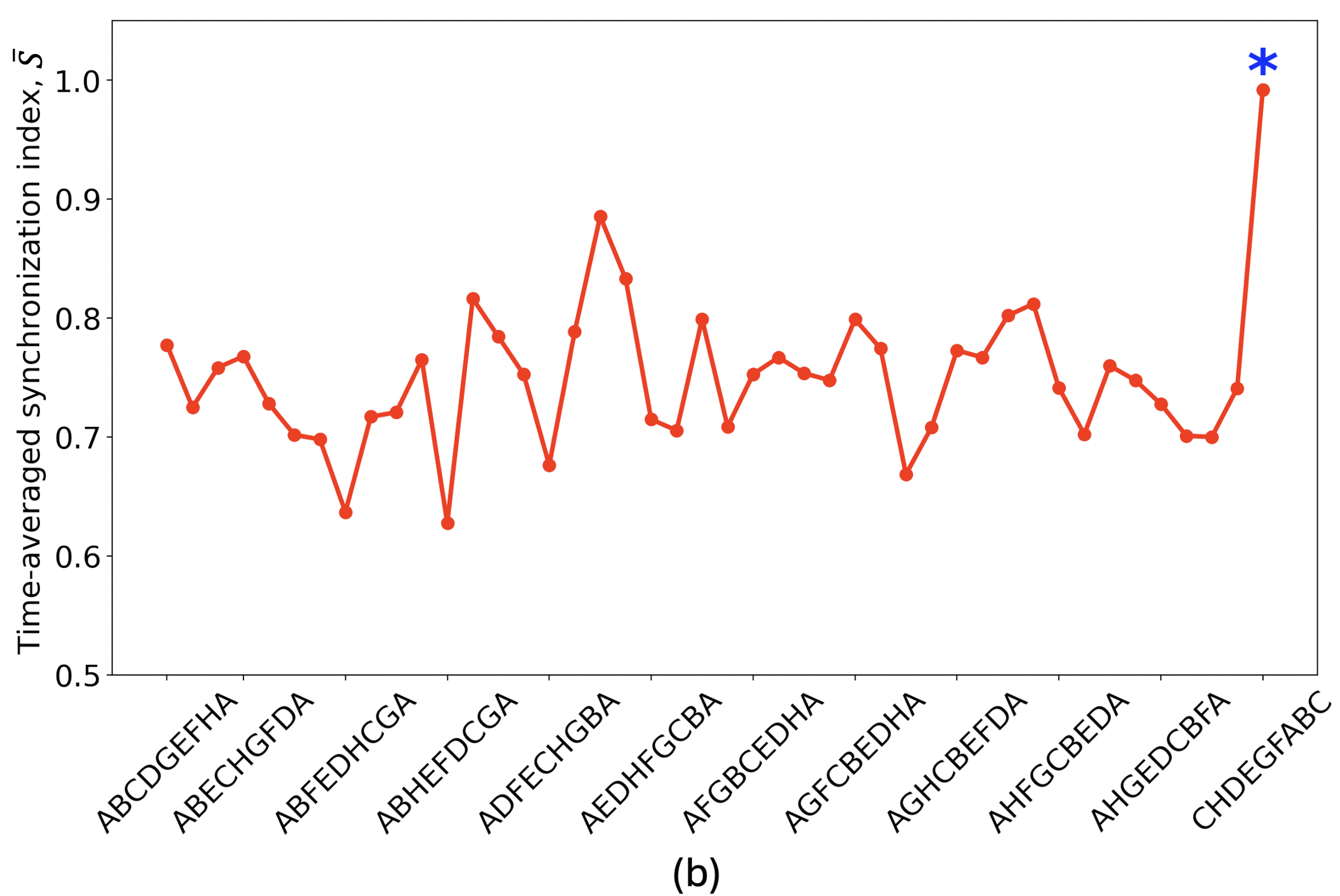}
    \caption{\textbf{The time-averaged synchronization index ($\overline{S}$) in the NESS is highest for the solution lanes of the \textit{Physarum}-selected solution compared to other distinct tours.}
(a) Histogram showing the distribution of tour lengths for eight-city TSP map with the time-averaged synchronization index ($\overline{S}$) values for all distinct valid tours in the NESS. The color of the bars represents the $\overline{S}$ values, averaged over all distinct tours for a specific tour length. Red markers indicate the standard deviation in $\overline{S}$ values across different tours of the same tour length. The bar marked with an asterisk represents tour length 128, indicating that among the many possible tours of this length, one was selected by \textit{Physarum}.
(b) $\overline{S}$ values in the NESS, plotted for all distinct tours of tour length 128, with the peak indicating the $\overline{S}$ value of the solution chosen by \textit{Physarum} (marked by the same asterisk). We list tours in lexicographic (dictionary) order by city label, from the first position to the last. Specifically, we first compare which city occupies the first position; among all tours sharing the same first city, we then compare the second city, and so on, until we have a complete ordering of all tours. The last tour in the list shows the \textit{Physarum}-selected solution (CHDEGFABC) displayed in Fig. 1e.}
\label{Tour_distribution}
\end{figure}


\begin{figure}[h]
    \centering
    \includegraphics[width=0.7\textwidth]{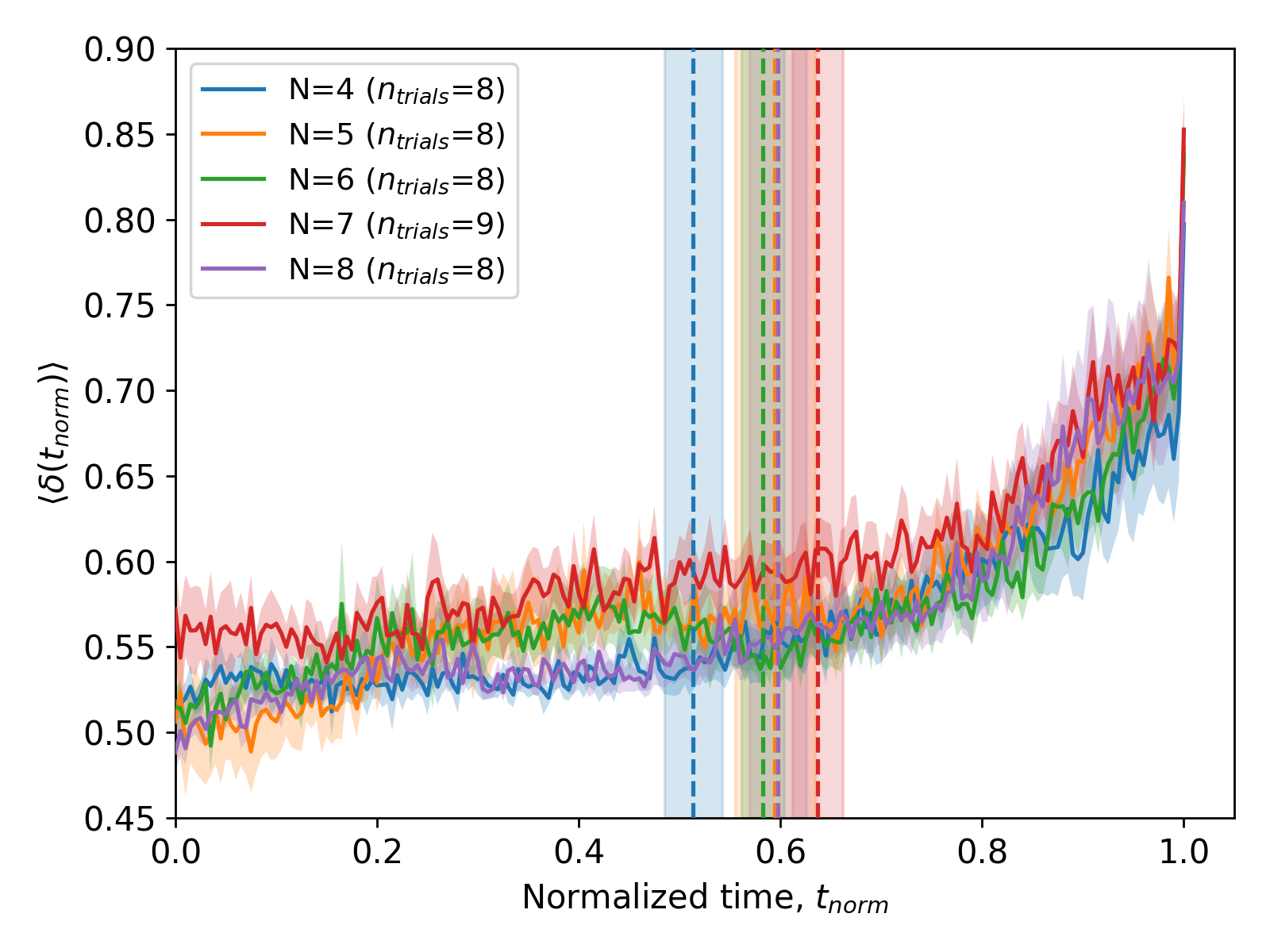}
    \caption{\textbf{The normalized synchronization ratio of solution lanes relative to non-solution lanes, $\delta(t)=\frac{S_{\text {Sol }}(t)}{S_{\text{Sol}}(t)+S_{\text {Non-Sol }}(t)}$, increases after the optical feedback is turned on and rises more rapidly once the system transitions to the non-equilibrium steady state (NESS)}. Curves represent the problem size-aggregated mean of this ratio, across all trials, with shaded regions indicating $\pm 1$ standard error. The time axis is normalized such that $t_{\text {norm }}=0$ corresponds to the onset of optical feedback and $t_{\text {norm }}=1$ corresponds to the time point at which the ratio reaches its maximum value for each trial. Vertical dashed lines indicate the problem size-aggregated mean transition time to the NESS, with shaded bands representing the standard error in the estimated NESS onset across trials for each problem size.}
    \label{Fig_Delta_Prime}
\end{figure}
 
At the start of the experiment, $S$ values for all lanes are similar and remain relatively constant in the preillumination region (Fig. ~\ref{TLSI}a). When the optical feedback is turned on, a subset of the lanes are illuminated. This illumination perturbs the dynamics of \textit{Physarum} branches, disrupting the synchronous phases across the branches. Consequently, the  $S$  values of non-solution lanes begin to decrease after successive illumination cycles. Although the solution lanes are not generally illuminated, the optical feedback control influences them indirectly through the central processing hub. As  $S$  values decrease in the non-solution lanes, the  $S$  values for the solution lanes either increase or remain constant near unity. In the NESS region, solution lanes achieve $S$  values close to the maximum ($\sim$ 1), while non-solution lanes reach a minimum $S$, exhibiting a noticeable and reproducible dip. This clear distinction between the two groups highlights the effect of an external electromagnetic drive on \textit{Physarum}'s intrinsic dynamics.

Next, we calculated the time-dependent synchronization index ($S(t)$ in Equation \ref{S}) of solution lanes for all distinct tours with the same tour length (128) as the solution chosen by \textit{Physarum}. The $S(t)$ value averaged over these tours (royal blue) is presented in Figure \ref{TLSI}b, with the standard deviation over these 44 tours displayed by the shaded transparent region. The $S(t)$ values for the solution lanes of the tour selected by \textit{Physarum} (regular blue) is shown again in Figure \ref{TLSI}b for comparison. On comparing the plots \ref{TLSI}a and  \ref{TLSI}b, we observe that the synchronization behavior of solution lanes averaged over all possible tours (Fig. \ref{TLSI}b, royal blue line) closely resembles that of the non-solution lanes in the \textit{Physarum}-selected solution (Fig. \ref{TLSI}a, orange line). This represents confirmation that \textit{Physarum} is dynamically optimizing over \textit{all} tours, not just those of a specific tour length value, because only the lanes that constitute the \textit{Physarum}-chosen solution (Figs. \ref{TLSI}a and b, blue line) exhibit the highest value of the synchronization index. This demonstrates that the \textit{Physarum}-selected solution corresponds to the state with the highest synchronization index among all competing tours of the same length, indicating that synchronization provides a direct criterion for solution selection.

\textcolor{black}{We also investigated the time-averaged synchronization indices ($\overline{S}$ in Equation \ref{S'}) for all distinct valid tours in the NESS region of the eight-city TSP map (Fig. \ref{fig:Scheme&chip}b). Although each experimental trial yields a single \textit{Physarum}-selected tour, we  evaluated $\overline{S}$ for every distinct valid tour by treating the lanes comprising each candidate tour as the solution-lane set.} We generated a histogram (Fig. \ref{Tour_distribution}a) displaying the distribution of tour lengths, with the number of distinct tours on the y-axis. Each bar represents the number of distinct tours for a specific tour length, while its color denotes the mean of the time-averaged synchronization indices, $\overline{S}$, computed across all distinct tours of that tour length in the NESS region. The bar marked with an asterisk corresponds to tour length 128, indicating that among the 44 possible solutions for this tour length, one was selected by \textit{Physarum}. This plot indicates that the time-averaged synchronization indices, averaged over all distinct tours for each particular tour length, are almost similar for all tours.

We then examined the time-averaged synchronization indices in the NESS region for the solution lanes of 44 distinct tours of length 128, as shown in Fig. \ref{Tour_distribution}b. The tours are arranged in lexicographic order with respect to city labels. We first list all tours whose first position is occupied by the city with the ``A'' label; within that group, we sort tours by the city in the second position, and so on for subsequent positions. After enumerating all tours for the first city in the first position, we move on to the next city in lexicographic label order and repeat this process until all tours are listed. (Note that any cyclic permutation of a given sequence of cities is equivalent, i.e., ABC = BCA = CAB, returning to the original city at the end.) The last tour in the list corresponds to the \textit{Physarum}-selected solution CHDEGFABC. 

For tour length 128, we enumerate only distinct tours by discarding cyclic permutations and fixing the first city in the first position for simplicity. Consequently, each listed tour (except for the \textit{Physarum}-selected solution) begins with city A. Due to our choice of lane labeling in Fig. 1e, the \textit{Physarum}-selected solution with the tour length of 128 begins with a different city.

From Figure \ref{Tour_distribution}b, we observe that the tour selected by \textit{Physarum} exhibits the highest time-averaged synchronization index value (marked by an asterisk). This further reinforces our argument that \textit{Physarum} consistently identifies solutions with the highest synchronization index. Similar trends in $\overline{S}$ values are observed across other tour lengths for $N=8$, as shown in Fig. S9 of the Supplementary Material.

\textcolor{black}{The synchronization dynamics observed for the representative $N=8$ trial with a tour length of 128 (Fig. \ref{TLSI}a) are consistently reproduced across other \textit{Physarum}-selected TSP solutions spanning multiple tour lengths and problem sizes $N=4-8$ (See Fig. S10,a-h in the Supplementary Material). Despite variations arising from distinct initial conditions before illumination in each trial, the temporal evolution of $S(t)$ remains qualitatively similar, indicating that the synchronization dynamics governing solution formation are robust to these variations. In three out of 41 trials, irregularities and dark budding regions appearing within the central processing hub of the \textit{Physarum} network substantially disturb the synchronization dynamics (Fig. S13 in the Supplementary Material). Nevertheless, \textit{Physarum }adapts to these internal perturbations and successfully identifies valid solutions even in these outlier cases.}

In order to quantitatively confirm that the enhanced synchronization of the solution lanes after the illumination starts is a consistent feature across all trials and problem sizes, we define the normalized synchronization ratio,
$\delta(t)=S_{\mathrm{Sol}}(t) /\left(S_{\mathrm{Sol}}(t)+S_{\text {Non-sol }}(t)\right)$. This quantity represents the weighted fraction for the solution lanes at time $t$ over the sum of synchronizations for the two lane groups. Values of $\delta(t)>0.5$ therefore indicate that the synchronization index of the solution lanes exceeds that of the non-solution lanes. The quantity $\delta(t)$ is computed for all trials, and the aggregated mean value, $\langle{\delta(t_{norm})}\rangle$, is obtained for each problem size $N=4-8$, as shown in Fig. \ref{Fig_Delta_Prime}. Thus, $\langle{\delta(t_{norm})}\rangle$ corresponds to the problem size-aggregated mean across all trials for each problem size. For each trial, the time axis is normalized such that $t_{norm}=0$ denotes the moment when the optical feedback pattern is turned on, while $t_{norm}=1$ corresponds to the time when the quantity $\delta(t)$ reaches its maximum value in the NESS region. The vertical dashed lines indicate the problem size-aggregated mean time at which the system transitions into the NESS. The shaded bands around these lines represent the standard error across multiple trials for each problem size in the estimated onset time of the NESS. The results shown here in Fig.~\ref{Fig_Delta_Prime} for $\langle{\delta(t_{norm})}\rangle$ are computed using a common normalized time grid of 200 points; corresponding results using a 500-point grid are provided in Fig.~S11a of the Supplementary Material, showing consistent behavior across grid resolutions.

As the illumination pattern is turned on  ($t_{norm}=0$), the value of $\langle{\delta(t_{norm})}\rangle$ begins to increase from near 0.5, indicating the gradual emergence of stronger synchronization in the solution lanes. A pronounced rise in $\langle{\delta(t_{norm})}\rangle$ is observed after the onset of the NESS, reaching a maximum at $t_{\mathrm{norm}}=1$ for each problem size. This confirms the earlier mentioned observation that the synchronization in the solution lanes becomes significantly stronger than non-solution lanes in this regime. The curves corresponding to different problem sizes remain broadly similar, with minor variations that likely arise from differences in active-lane geometry and \textit{Physarum}'s initial conditions. 
We also compute a related ratio $\eta(t)=\frac{\left|S _{\mathrm{Sol}}(t)-S_{\mathrm{Non}-\mathrm{Sol}}(t)\right|}{S_{\mathrm{Sol}}(t)+S_{\mathrm{Non}-\mathrm{Sol}}(t)}$, for all trials. The aggregated mean values $\langle{\eta(t_{norm})}\rangle$ for each problem size are shown in Fig.~S11b-c of the Supplementary Material for both 200- and 500-point normalized time grids.
This quantity shows qualitatively similar behavior to $\langle{\delta(t_{norm})}\rangle$, increasing after the onset of illumination and exhibiting a faster rise once the system reaches the steady state, before reaching a maximum at $t_{norm}=1$. 
\textcolor{black}{Additionally, to illustrate the distribution of the synchronization ratio across the trials, we computed time-averaged $\bar{\delta}$ by averaging $\delta(t)$ over the pre-illumination interval, the first and second halves of the optical feedback interval, and the NESS interval for all 41 experimental trials. The corresponding violin plots are shown in Fig. S12 of the Supplementary Material. Consistent with Fig. ~\ref{Fig_Delta_Prime}, the distributions shift progressively toward higher $\bar{\delta}$ values from the first to the second half of the optical feedback interval, reaching their maximum in the NESS.}

Overall, these results demonstrate the statistical robustness of the stronger synchronization in solution lanes compared to the non-solution lanes during the solution search across the 41 TSP trials performed for different problem sizes.
Since the behavior of the synchronization index remains consistent in tracking and distinguishing \textit{Physarum} states across various problem sizes and tour lengths, it  serves as a good figure of merit to identify and predict the selected solution. It can potentially be used to coax \textit{Physarum} toward more optimal TSP solutions, in a more efficient manner than our previous implementations. 
\textcolor{black}{
\subsection{Early prediction of solution lanes from synchronization dynamics}
In the previous subsection, we showed that the synchronization indices of the solution and non-solution lanes begin to bifurcate shortly after the optical feedback is activated, indicating the emergence of the lane group that will ultimately constitute \textit{Physarum}'s solution to the TSP instance. Here, we quantitatively evaluate the capability of the synchronization metrics to predict the final solution across all 41 experimental trials spanning problem sizes $N=4-8$. For each experimental trial, we computed the synchronization index for each of the $N!$ possible tour permutations corresponding to the TSP instance of problem size $N$. To facilitate aggregation across trials with different TSP solution search times, we normalized the optical-feedback interval by introducing the normalized time variable
\begin{equation}
\tau_1=\frac{t-t_{\text{illumination}}}{t_{\text {NESS}}-t_{\text {illumination}}},
\end{equation}
where $t_{\text {illumination}}$ denotes the onset of optical feedback and $t_{\mathrm{NESS}}$ is the time at which the system reaches the NESS. Next, we quantified the similarity between the \textit{Physarum}-selected solution and every valid alternative tour by calculating the fraction of solution lanes shared between the two tours. The similarity score $r$ is therefore defined as
\begin{equation}
r=\frac{\text {Number of lanes shared with the \textit{Physarum}}\text{-selected solution}}{N},
\end{equation}
where $N$ is the number of solution lanes. For each experimental trial corresponding to a problem size $N$, the similarity score was computed for each of the $N!$ valid tour permutations, and the corresponding synchronization index ($S$) was assigned to each tour.
The $S$ values associated with the same similarity score $r$ were then averaged to obtain the similarity score-aggregated mean synchronization index for the $j$-th trial
\begin{equation}
S_j^{(r)}=\frac{1}{n_r} \sum_{i \in r} S_{ij}
\end{equation}
where $n_r$ is the number of valid tours with similarity score $r$, and the summation extends over all tours belonging to that similarity group. Thus, the similarity score-aggregated mean synchronization index, $S_j^{(r)}$, was computed for every trial corresponding to a given problem size. The problem size-aggregated mean synchronization index was then obtained by averaging $S_j^{(r)}$ across all trials of the same problem size $N$ for a given similarity score,
\begin{equation}
\left\langle S^{(r)}\right\rangle_N=\frac{1}{n_{\text {trials }}} \sum_{j=1}^{n_{\text {trials }}} S_j^{(r)}
\end{equation}
where $n_{\text {trials }}$ denotes the total number of trials for the problem size $N$.
}

\begin{figure}
    \centering
    \includegraphics[width=0.9\linewidth]{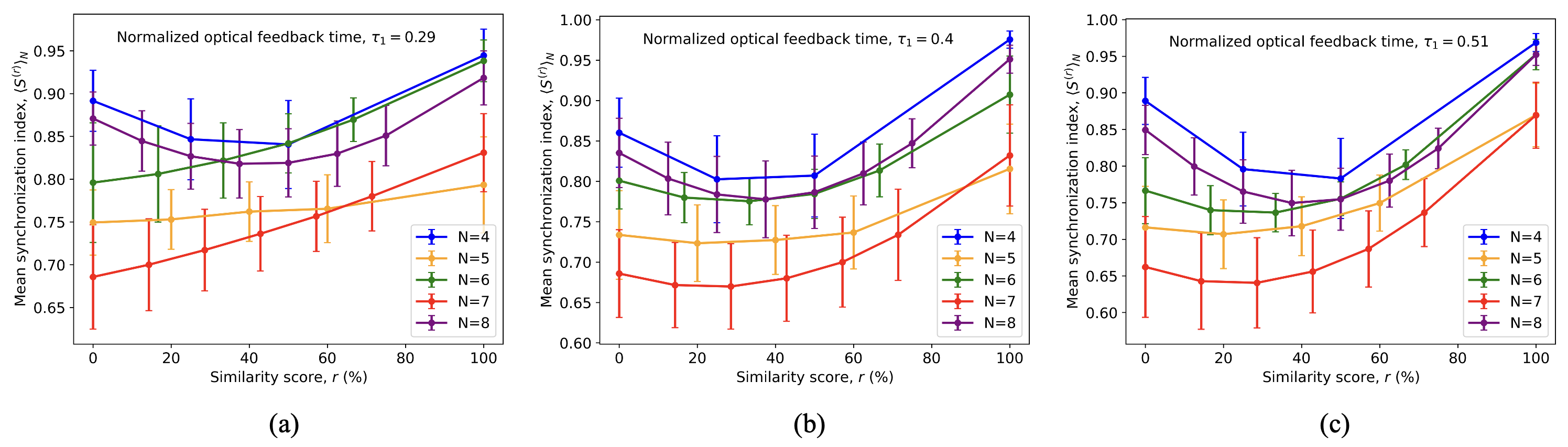}
    \includegraphics[width=0.7\linewidth]{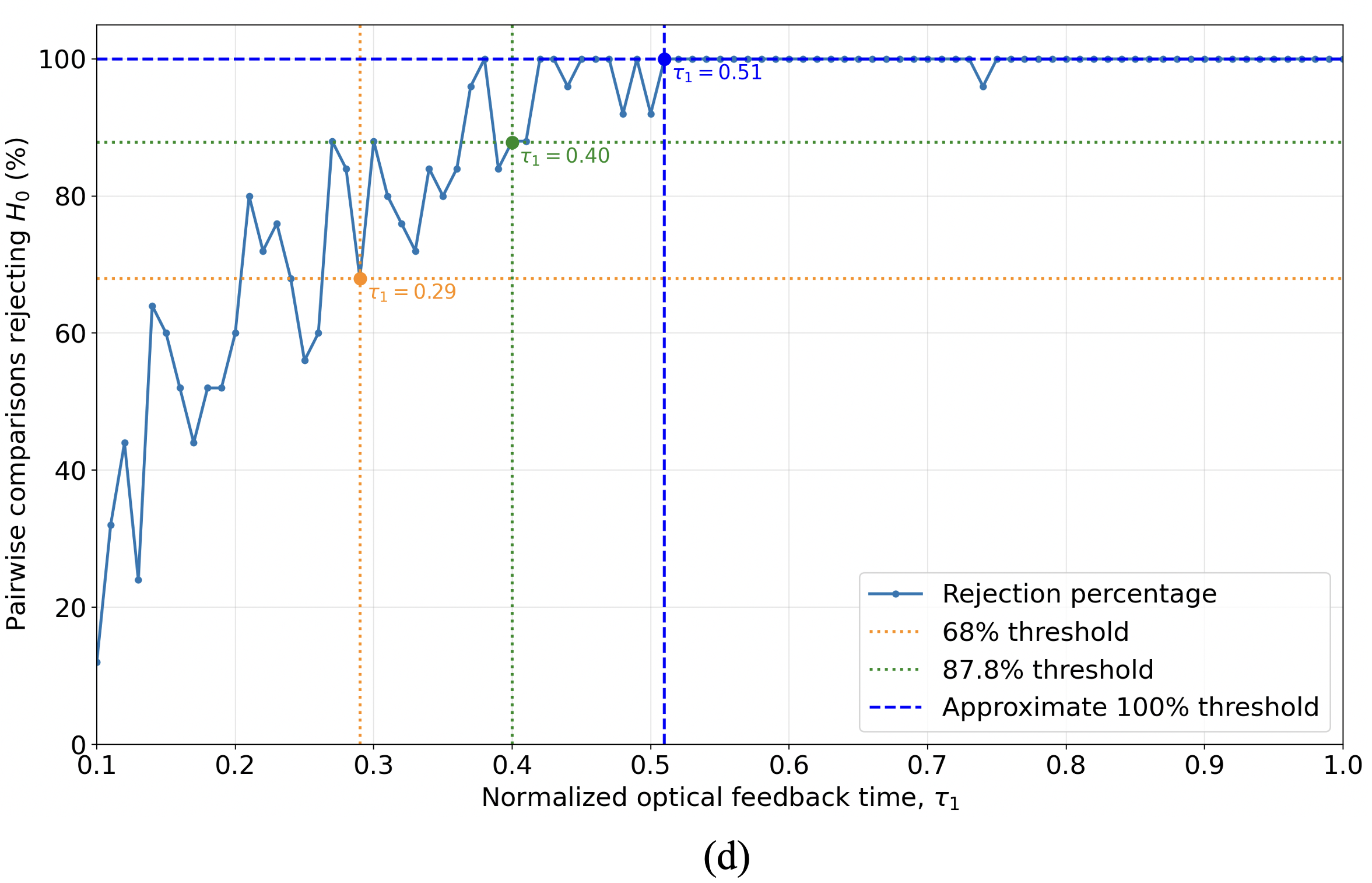}
    \caption{\textcolor{black}{\textbf{Synchronization-based metrics reliably predict the \textit{Physarum}-selected solution with statistical significance within the first half of the optical-feedback interval, with approximately 68\% of pairwise comparisons reaching significance by one-third of the optical-feedback interval and nearly 100\% by its midpoint, well before the onset of the NESS.} The figure shows problem size-aggregated mean synchronization index, $\left\langle S^{(r)}\right\rangle_N$, as a function of similarity score for normalized optical-feedback times (a) $\tau_1=0.29$, (b) $\tau_1=0.40$, and (c) $\tau_1=0.51$. The problem size-aggregated mean synchronization index is obtained by averaging over trials of the same problem size ($N=4$--8), and the error bars represent $\pm1$ standard error. (d) Percentage of pairwise comparisons rejecting the null hypothesis ($p<0.05$) in the paired $t$-test between the \textit{Physarum}-selected solution (100\% similarity) and lower-similarity tours as a function of normalized optical-feedback time, $\tau_1$. The vertical dashed lines indicate the optical-feedback times shown in panels (a--c), corresponding to approximately 68\%, 87.8\%, and nearly 100\% of statistically significant pairwise comparisons, respectively.}}
    \label{fig:Sim_Score}
\end{figure}

\textcolor{black}{The problem size-aggregated mean synchronization index, $\left\langle S^{(r)}\right\rangle_N$, as a function of similarity score is shown in Fig. ~\ref{fig:Sim_Score}a-c for three representative normalized optical-feedback times, $\tau_1$= 0.29, 0.40, and 0.51. These time points were selected based on the statistical analysis presented below and correspond to distinct stages in the emergence of predictive synchronization. Error bars represent $\pm 1$ standard error across trials for each problem size. At all three normalized times, the \textit{Physarum}-selected solution, defined as the unique tour with a similarity score of $100 \%$, exhibits the highest mean synchronization index. The separation between the $100 \%$ \textit{Physarum}-selected group and alternate tours with lower similarity scores is already evident at $\tau_1=0.29$ and becomes progressively more pronounced at $\tau_1=0.40$ and 0.51, indicating that synchronization increasingly discriminates the final solution from competing tours as the optical feedback evolves.}

\textcolor{black}{To determine whether the observed separation between the \textit{Physarum}-selected solution ($r= 100 \%$) and tours with lower similarity scores ($r<100 \%$) is statistically significant, we performed paired $t$-tests. A paired test was appropriate because we examined the difference
\begin{equation}
\Delta S_N^{(r)}=\left\langle S^{(100 \%)}\right\rangle_N-\left\langle S^{(r<100 \%)}\right\rangle_N
\end{equation}
where each comparison involves a paired set of problem size-aggregate mean synchronization indices at a given normalized optical-feedback time. Before applying the paired $t$-test, we verified the normality of $\Delta S_N^{(r)}$ using the Shapiro-Wilk test. For the paired $t$- test, the null hypothesis ($H_0$) states that no difference exists between the mean synchronization indices of the \textit{Physarum}-selected solution ($r=100 \%$) and tours with lower similarity scores ($r<100 \%$), whereas the alternative hypothesis ($H_1$) states that the mean synchronization index of the \textit{Physarum}-selected solution is greater than that of tours with lower similarity scores.}

\textcolor{black}{The paired $t$-tests were performed for normalized optical-feedback times ranging from $\tau=0.10$ to 1.00 in increments of 0.01. The resulting $p$-value heat maps as a function of normalized optical-feedback time and similarity score are presented in Fig. S14 of the Supplementary Material. For each problem size $N$, the problem-size-aggregated mean synchronization index of the \textit{Physarum}-selected solution ($100{\%}$ similarity group) was compared with those of the remaining $N-1$ lower-similarity groups. Across $N=4-8$, this resulted in a total of 25 pairwise comparisons at each normalized time point. The percentage of pairwise comparisons rejecting the null hypothesis ($H_0$) as a function of normalized optical-feedback time is shown in Fig. ~\ref{fig:Sim_Score}d. At $\tau_1=0.29,68 \%$ of the comparisons rejected the null hypothesis ($p<$ 0.05). This fraction increased to $87.8 \%$ at $\tau_1=0.40$ and reached ${\sim}100\%$ at $\tau_1=0.51$. Thus, by approximately the midpoint of the optical-feedback interval, the synchronization indices of the \textit{Physarum}-selected solutions are statistically distinguishable from those of every lower-similarity score tour group, demonstrating robust early prediction of the solution lanes. Even more so, at less than a third of the optical-feedback interval, the \textit{Physarum}-selected solutions have distinguished themselves at the level of $\pm1\sigma$ for the normal distribution of pairwise comparisons. These results provide a quantitative measure of the predictive power of synchronization and demonstrate that the lanes ultimately forming the \textit{Physarum}-selected solution can be identified with high statistical confidence when the system has progressed through only approximately $30 \%$ of the optical-feedback interval.}

\textcolor{black}{To compare the predictive capability of the synchronization metric with that of the lane-state variable $X_{V k}$, we next examined whether $X_{V k}$-based statistics could identify the final \textit{Physarum}-selected solution during the optical-feedback interval. Two possible prediction criteria were considered. First, the standard deviation of the $X_{V k}$ values in the solution lanes, $\sigma_{X_{V k}}^{\text {sol }}$, should decrease substantially in the optical feedback interval if the solution lanes become distinguishable. Alternatively, the standard deviation of the $X_{V k}$ values for the solution lanes should become sufficiently smaller than that for the non-solution lanes, thereby allowing the solution-forming lane group to be distinguished reliably.}

\textcolor{black}{To test these hypotheses, we computed the standard deviations of the $X_{V k}$ values for the solution and non-solution lanes, denoted by $\sigma_{X_{V k}}^{\text {sol }}$ and $\sigma_{X_{V k}}^{\text {non-sol }}$, respectively. The resulting values were averaged across trials for each problem size and analyzed as functions of both the normalized optical-feedback time ($\tau_1$), and the normalized time referenced to the minimum $X_{Vk}$ dispersion in NESS,
\begin{equation}
\tau_2=\frac{t-t_{\text {illumination}}}{t_{\sigma_{\min }}-t_{\text {illumination}}}.
\end{equation}
Here, $t_{\sigma_{\min}}$ denotes the time at which $\sigma_{X_{V k}}$ reaches its minimum value in the NESS. The corresponding results are presented in Fig. S15a-b of the Supplementary Material.}

\textcolor{black}{The $X_{V k}$ dispersion analysis reveals that neither criterion provides reliable early prediction of the final solution. As shown in Fig. S15a of the Supplementary Material, the standard deviation of the $X_{V k}$ values in the solution lanes remains relatively large (approximately $0.15-0.31$) throughout the optical-feedback interval and decreases substantially only as the system approaches the NESS. Furthermore, Fig. S15b of the Supplementary Material demonstrates that, during the optical-feedback interval, the standard deviation of the solution lanes remains consistently higher than that of the non-solution lanes (approximately $0.08-0.14$). Thus, the $X_{V k}$ values in the solution lanes remain highly variable throughout the optical-feedback interval and do not exhibit the low dispersion required for reliable identification of the final solution. Consequently, $X_{Vk}$-based metrics cannot reliably predict the solution before the onset of the NESS.}

\textcolor{black}{In contrast, the synchronization metric exhibits statistically significant separation between the \textit{Physarum}-selected solution and competing tours after only approximately $30 \%$ of the optical-feedback interval has elapsed, with nearly $68 \%$ of the pairwise comparisons rejecting the null hypothesis at $\tau_1=0.29,87.8 \%$ at $\tau_1=0.40$, and $100 \%$ at $\tau_1=0.51$. These results demonstrate that synchronization provides a substantially more sensitive indicator of the emerging solution than $X_{Vk}$-based metrics, enabling reliable prediction of the solution-forming lanes well before the system reaches the NESS.}

\begin{figure}
    \centering
    \includegraphics[width=1\linewidth]{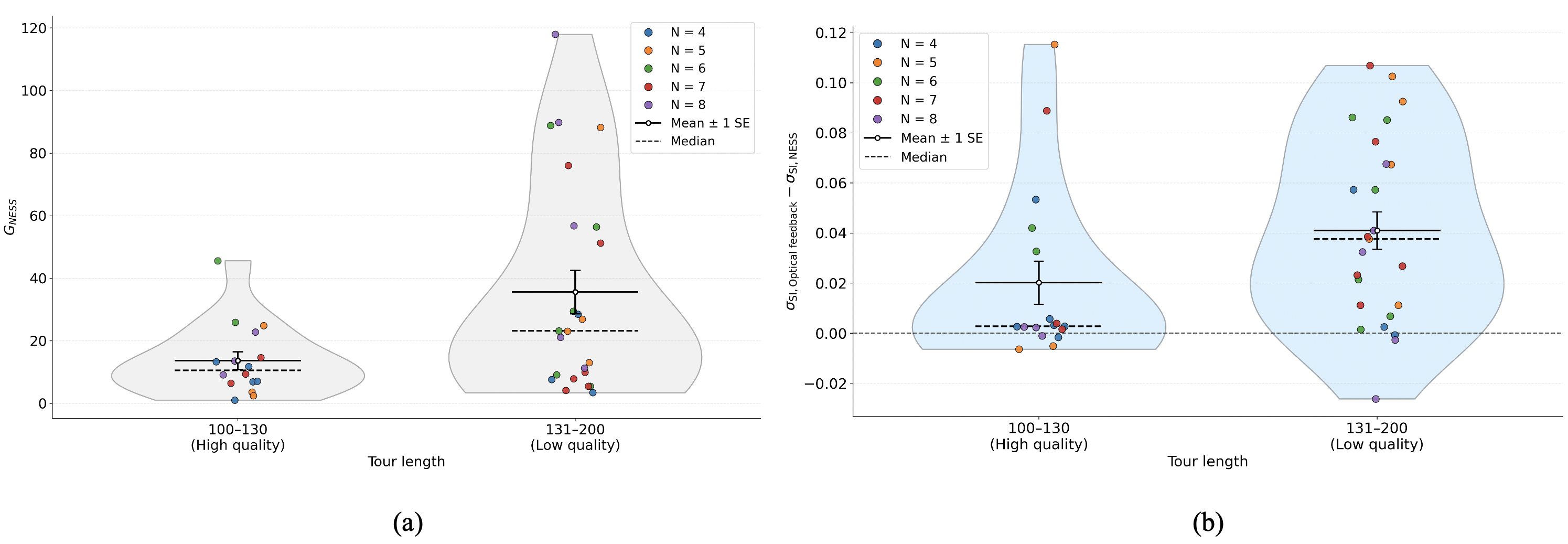}
    \caption{\textcolor{black}{\textbf{Power spectrum and synchronization are strongly correlated with the quality of the \textit{Physarum}-selected TSP solutions, with higher-quality solutions exhibiting lower amplification gain and earlier stabilization of the synchronization dynamics.}
    The figure shows (a) amplification gain in the NESS, $G_{\mathrm{NESS}}$, defined as the ratio of the mean low-frequency (4--50 mHz) PSD peak of the solution lanes to that of the non-solution lanes, for high-quality (tour length $100$--$130$) and low-quality (tour length $131$--$200$) \textit{Physarum}-selected solutions. (b) Reduction in the temporal dispersion of the synchronization index between the optical-feedback and NESS regions, $\Delta \sigma_{S}^{\text {Opt.feed.→NESS }}=\sigma_{S}^{\mathrm{Opt.feed.}}-\sigma_{S}^{\text{NESS}}$, shown for the same solution-quality groups. Here, $\sigma_{S}^{\mathrm{Opt.feed.}}$ and $\sigma_{S}^{\mathrm{NESS}}$ denote the temporal standard deviations of the synchronization indices of the solution lanes in the optical-feedback and NESS regions, respectively. The width of each violin represents the distribution density of the data. Colored circles denote individual trials for problem sizes $N=4-8$. The solid horizontal line and whiskers represent the mean $\pm1$ standard error, whereas the dashed horizontal line denotes the median.}}
    \label{fig:Solution_quality_violin}
\end{figure}

\textcolor{black}{
\subsection{Correlation between spectral dynamics, synchronization dynamics, and solution quality}}
\textcolor{black}{Having demonstrated the predictive capability of synchronization, we next examine how the spectral and synchronization dynamics are related to the quality of the \textit{Physarum} TSP solutions. The complete dataset comprises 41 experimental trials spanning problem sizes $N=4-8$. As described in the Introduction, the TSP instances were constructed such that the shortest and longest valid tours have lengths of 100 and 200, respectively. Based on the criterion established in the Introduction, tours with lengths less than or equal to 130 are classified as high-quality solutions. Accordingly, 17 of the 41 \textit{Physarum}-selected tours fall into the high-quality category, whereas the remaining 24 tours, with lengths greater than 130, are classified as low-quality solutions.}

\textcolor{black}{To investigate how the spectral and synchronization dynamics are related to solution quality, we evaluated two dynamical quantities for each trial and grouped the results according to the quality of the corresponding \textit{Physarum}-selected solution. Specifically, we considered: (a) the amplification gain, $G_{\mathrm{NESS}}$, defined as the ratio of the mean PSD of the solution lanes to that of the non-solution lanes in the low-frequency region $(4-50\ \mathrm{mHz})$; and (b) the temporal standard deviation of the synchronization indices of the solution lanes during the optical-feedback and NESS regions. }

\textcolor{black}{To quantify the temporal stabilization of the synchronization dynamics, we computed the reduction in the temporal dispersion of the synchronization indices between these two regions,
\begin{equation}
\Delta \sigma_{{S}}^{\text {Opt.Feed. } \rightarrow \mathrm{NESS}}=\sigma_{S}^{\text {Opt.Feed. }}-\sigma_{S}^{\mathrm{NESS}}
\end{equation}
where $\sigma_{S}^{\text {Opt.Feed.}}$ and $\sigma_{S}^{\text {NESS }}$ denote the temporal standard deviations of the synchronization indices of the solution lanes during the optical-feedback and NESS regions, respectively. Larger values of $\Delta \sigma_{S}^{\text {Opt.Feed.} \rightarrow \text {NESS}}$ indicate a greater reduction in synchronization dispersion in the NESS. The distributions of $G_{\text {NESS}}$ and $\Delta \sigma_{S}^{\text {Opt.Feed.} \rightarrow \text {NESS}}$ for high- and low-quality solutions are shown in Fig.~\ref{fig:Solution_quality_violin} using violin plots. The width of each violin represents the distribution density of the data. The solid horizontal line indicates the mean, with whiskers denoting $\pm 1$ standard error, whereas the dashed horizontal line indicates the median. Data from different problem sizes are shown using different colors.}

\textcolor{black}{For the amplification gain, $G_{\text{NESS}}$, both the mean and median are substantially lower for high-quality solutions (tour length $\leq 130$) than for low-quality solutions (tour length$>130$), as shown in Fig.~\ref{fig:Solution_quality_violin}a. The mean amplification gain for high-quality solutions is ${\sim}15$, with the median lying slightly below the mean. In contrast, the mean amplification gain for low-quality solutions increases to ${\sim} 36$, while the median is slightly over 20. These results demonstrate that high-quality solutions are generally associated with lower amplification gains than low-quality solutions, suggesting that high-quality solutions stabilize earlier, thereby limiting the subsequent growth of the amplification gain.}

\textcolor{black}{Similarly, the reduction in the temporal standard deviation of the synchronization indices between the optical-feedback and NESS regions, $\Delta \sigma_{S}^{\text {Opt.Feed.} \rightarrow \text {NESS}}$, is shown in Fig.~\ref{fig:Solution_quality_violin}b. High-quality solutions exhibit a mean value of 0.02, with the median lying below the mean and remaining close to zero. In contrast, both the mean and median for low-quality solutions are close to ${\sim} 0.04$. The smaller value of $\Delta \sigma_{S}^{\text {Opt.Feed.} \rightarrow \text {NESS}}$ for high-quality solutions indicates that synchronization has already become largely stabilized during the optical-feedback phase, requiring only a modest reduction in temporal dispersion before reaching the NESS. Conversely, the larger values observed for low-quality solutions suggest that synchronization continues to evolve throughout the optical-feedback interval and only stabilizes upon reaching the NESS. Consequently, low-quality solutions exhibit a greater reduction in synchronization dispersion between the optical-feedback and NESS regimes. Collectively, these analyses demonstrate that solution quality is strongly correlated with both the spectral and synchronization dynamics of the \textit{Physarum} network. These findings suggest that theoretical improvements to the Hopfield neural network and experimental optimizations of the optical-feedback protocol, aimed at promoting favorable spectral and synchronization dynamics, may further improve the quality of the solutions generated by \textit{Physarum}-machine interfaces.}

\section{Discussion}

By implementing an optical feedback loop based on a modified Hopfield neural network, we analyzed the behavior of \textit{Physarum} branches within a stellate TSP chip, distinguishing between solution and non-solution lanes. The separation of these groups under illumination highlights the problem-solving mechanism of \textit{Physarum}. Upon reaching a solution, the system stabilizes in a non-equilibrium steady state (NESS), where the synchronization index ($S$ in Equation \ref{S}) of solution lanes achieves its maximum value, while non-solution lanes exhibit a minimum. This  bifurcation in synchronization indices indicates that a higher synchronization metric is a defining trait of the solution and can thus be used to identify and predict the solution selected by \textit{Physarum} for the TSP. \textcolor{black}{ By evaluating the synchronization indices of every valid TSP tour as a candidate solution for each of the 41 experimental trials spanning problem sizes $N=4-8$, we demonstrate that the \textit{Physarum}-selected solution consistently exhibits the highest synchronization index. Moreover, the difference between the synchronization index of the \textit{Physarum}-selected solution and those of all alternative tours becomes statistically significant within the first half of the optical-feedback interval, enabling reliable prediction of the final solution well before the system reaches the NESS.}


A notable feature of the organismal dynamics is that this bifurcation of $S$ values between solution and non-solution lanes occurs well before the onset of NESS, occurring early in the optical feedback region. This underscores how quickly \textit{Physarum} adapts to the optical feedback, effectively arriving at a valid TSP solution in advance of reaching the NESS. A more controlled study of exactly when the \textit{Physarum} state distinguishes a valid TSP solution is planned, which would potentially change our estimates of the scaling behavior of \textit{Physarum} solution time with problem size.

Another intriguing observation comes from the behavior of $S(t)$ for solution lanes in the pre-illumination region (Figs. 7 and 8). In this phase, optical feedback has not yet been activated, and the problem defined by the modified Hopfield weight matrix has not yet been presented to \textit{Physarum} through the updating illumination pattern. Still, in some cases, we observe an increase in the $S(t)$ values of the solution lanes. This suggests that \textit{Physarum} branches are not merely extending within the chip’s lanes but are actively adapting to the chip environment. Such adaptation may reflect a form of preparatory self-organization, where \textit{Physarum} acclimatizes to the chip but effectively ‘initializes’ to solve the problem, even before the external light stimulus is applied. Such an astounding property would indeed require further evidence and controlled verification to accept.

The slime mold's intricate branch configurations are governed by the optical feedback and mediated by the central \textit{Physarum} hub, which acts as a computing, processing, and integration unit exhibiting highly nonlinear dynamics. \textit{Physarum} thus demonstrates traits of reservoir computers \cite{tanaka2019recent,nakajima2020physical}, which leverage the natural dynamics of physical systems to serve as computational devices. Unlike traditional neural networks, which require intensive parameter tuning across all layers, reservoir computers confine training to the readout layer, resulting in faster and more stable learning outcomes. Various dynamical systems, including biological substrates such as \textit{Physarum}, can be exploited as reservoirs for computing with the appropriate probe/stimulus and readout, as long as they provide reproducible outputs for similar input conditions. We have demonstrated this type of reproducibility in our amoeba-based biocomputer. 


\textcolor{black}{The present study complements a companion work by two of the authors (SB, PK) and coworkers on the physical limits of computation in \textit{Physarum}~\cite{bajpai2025morphological}. There, we derive distinct upper bounds on \textit{Physarum}’s computational capacity from its hydrodynamical, chemical (ATP), kinetic, and quantum-optical degrees of freedom, each obtained from area and perimeter time-series data explored by the organism during morphological growth in an agar arena. Likewise, from the explored area in the chip, the chemical and quantum-optical bounds on the number of logical operations for the $N^2$ branches can thus be estimated. Additionally, by using the lane perimeter and the characteristic \textit{Physarum} speed, the hydrodynamical and kinetic-energy bounds can be computed. Considering the eight-city chip used in the experiments, and applying the formulations of these bounds derived in the companion work~\cite{bajpai2025morphological}, we roughly estimate the upper limits on the number of operations that can be performed by both the central hub and by each lane over a 24-hour period. The resulting upper bounds on \textit{Physarum}’s computational capacity, derived from the different physical degrees of freedom described above, are summarized in Table S4 of the Supplementary Material. \textit{Physarum} in the central circular hub is estimated to perform no more than $5.20 \times 10^{30}$ operations per day, while \textit{Physarum} in a single lane is estimated to perform no more than $5.71 \times 10^{28}$ operations per day. Together, these branch- and hub-level bounds provide a quantitative framework for understanding and potentially enhancing the solution quality and computational efficiency of \textit{Physarum} in solving the traveling salesperson problem.}


\subsection*{\textcolor{black}{Conclusion and Outlook}}
\addcontentsline{toc}{subsection}{Conclusion and Outlook}

\textcolor{black}{Whereas previous studies primarily demonstrated \textit{Physarum}’s ability to solve the TSP and identify high-quality solutions in the non-equilibrium steady state (NESS) \cite{zhu2018remarkable, zhu2013amoeba}, the present work establishes a quantitative framework for characterizing the emergence of the selected solution, predicting it shortly after illumination, and uncovering the collective synchronization dynamics that govern the computation.  We show that \textit{Physarum} exhibits distinct synchronization behavior between solution and non-solution lanes. By analyzing \textit{Physarum}’s occupation in different lanes, we calculated the lane states ($X_{Vk}$), which revealed a higher degree of synchrony in the phases of solution lanes during the NESS, whereas the phases of non-solution lanes exhibited substantially lower synchrony.}

To quantify this behavior, we calculated an order parameter known as the time-dependent synchronization index ($S(t)$), representing the narrowing of the standard deviation in phases among a group of pseudopod oscillators, with each lane considered an individual oscillator. Using the Hilbert transform, we determined the phases of the individual oscillators and subsequently computed the time-dependent synchronization indices for both solution and non-solution lanes. We plotted the $S(t)$  values over time and identified three distinct regions. In the pre-illumination region, all lanes behaved similarly, with $S(t)$ values close to each other. However, once the illumination was activated by the modified Hopfield neural network at threshold values of $X_{Vk}$, the  $S(t)$ of the solution lanes remained relatively constant, while the $S(t)$ of the non-solution lanes decreased due to perturbation by the applied light. This separation became more pronounced in the NESS region, where the $S(t)$ of solution lanes approached the maximum value (near unity), and the $S(t)$ of non-solution lanes reached a minimum. This trend is consistently observed in \textit{Physarum} TSP solutions across various tour lengths, for problem sizes ranging from $N=4-8$.
 
This distinct difference in time-dependent synchronization behavior between solution and non-solution lanes highlights the utility of $\delta(t)=S_{\text {Sol }}(t) /\left(S_{\text {Sol }}(t)+S_{\text {Non-sol }}(t)\right)$ as a good figure of merit for identifying and predicting the solution and for guiding \textit{Physarum} toward better solutions, potentially with reduced algorithmic overhead (i.e., fewer iterations of the optical feedback controlled by the modified Hopfield network). \textcolor{black}{By effectively tuning the synchronization indices, it may be possible to drive \textit{Physarum} toward solutions that are closer to optimal, further solidifying its role as an organism capable of finding high-quality approximate solutions to complex optimization problems such as the TSP, without a brain or neurons.}

We also demonstrated that for a small number of cities, \textit{Physarum}  exhibits a linear dependence of computation time on the number of cities. In contrast, highly efficient silicon-based algorithms, such as the Lin-Kernighan-Helsgaun (LKH) heuristic, show at best a quadratic scaling with problem size. Additionally, as highlighted in Fig. S7 of the Supplementary Material, the \textit{Physarum}-inspired Improved AmoebaTSP model exhibits slightly better scaling than the LKH heuristic in both iteration count ($\sim \sqrt{N}$) and computation time for problem sizes in the range $N=10$ to $N=100$. Such scalings suggest the presence of a powerful form of noise-enhanced parallel processing and dynamical global optimization within \textit{\textit{Physarum}’s} body, highlighting its exceptional problem-solving capabilities.

While the classical Hopfield model is powerful for small-scale problems, it suffers from certain limitations such as low storage capacity, susceptibility to local minima, and poor scalability. The quantum Hopfield model naturally addresses these challenges by leveraging quantum principles like superposition, entanglement, and tunneling, enabling enhanced computational power and optimization capabilities. \textcolor{black}{More details on the quantum Hopfield model are discussed in Section S11 of the Supplementary Material.}

In addition to improving the optical feedback control, we are planning to implement an improved Hopfield model for the neural network so that \textit{Physarum} can search for an optimal solution in shorter time. Many quantum-inspired computational approaches have been developed to better simulate complex biological phenomena with clear experimental observables, including relevant quantum degrees of freedom to explain the highly efficient rates of enzyme catalysis \cite{gori2023second}, protein folding \cite{unke2024biomolecular}, long-range biomolecular organization and targeting \cite{nardecchia2018out,lechelon2022experimental, azizi2023examining, Faraji2025}, cytoskeletal signaling \cite{babcock2024ultraviolet, Kurian2025}, and anomalous protein aggregation in various degenerative pathologies \cite{patwa2024quantum}. We expect that probing such photoexcited quantum degrees of freedom with specific excitation wavelengths and at shorter time intervals may improve \textit{Physarum}'s dynamic search and optimization capability across a highly complicated free energy landscape.

\section{Materials and Methods}

\subsection{Experimental} 

\subsubsection{Preparation of \textit{Physarum} sample}

A plasmodium of the slime mold \textit{\textit{Physarum} polycephalum} \textcolor{black}{(cross strain HU195$\times$200, obtained from Hokkaido University)} was placed on a $1\%$ agar plate with oat flakes at 25°C in the dark. The nutrient-rich agar used for the experiments was prepared using the following components: Ultra-pure water: 100 mL, Bacto-agar: 1.5 g, Glucose: 0.36 g, KCl: 0.074 g, Malt extract: 1 g and Peptone: 0.1333 g. To prevent moisture evaporation, the surface of the plate was covered with a plastic layer.

\subsubsection{Stellate chip fabrication}

The stellate chip used in the experiment was fabricated from ultra-thick photoresist resin through photolithography. The chip had the following dimensions: a thickness of approximately 0.1 mm, a total diameter of ~23.5 mm, a central disk diameter of ~12.5 mm, and lane dimensions of 3 mm in length and 0.45 mm in width.

\subsubsection{Experimental setup}

The experiments were conducted in a dark thermostat and humidostat chamber maintained at 28 ± 0.5°C and 70 ± 5\% humidity. The sample was illuminated from the top using a projector emitting white light with intensity of 352 µW/mm². To confine the organism within the stellate chip, the outer edge of the chip was continuously illuminated to prevent \textit{Physarum} from moving beyond the boundary. Real-time image processing was performed using custom-written codes, and time-lapse video images were captured with a video camera at intervals of 6 seconds. The darkness of each pixel in the recorded images corresponded to the vertical thickness of the organism at the respective site.

\subsection{Image processing and extraction of the lane state $X_{Vk}$}
Images extracted from the compressed raw experimental videos were converted to grayscale and binarized using Otsu's thresholding method \cite{sezgin2004survey} applied to each frame. A virtual mask was constructed consisting of a black background with a white circle of radius 155 pixels and radial lanes arranged along the circumference to reproduce the geometry of the stellate chip. The trapezoidal lanes of the mask had a length of 90 pixels, an inner width of 7 pixels (near the circle) and an outer width of 12 pixels (away from the circle). Prior to the experiment, a reference image of the empty chip (without \textit{Physarum}) was acquired and combined with the virtual mask through pixel-wise logical AND operations to define 64 distinct regions of interest (ROIs), each corresponding to an individual lane on the chip and smaller than its trapezoidal lane mask. Each ROI thus contained a single white rectangular lane of the chip on a black background (pixel value=0). The total area of a lane without \textit{Physarum} was determined as the number of white pixels (pixel value=255) in each ROI. These ROIs from the trial-specific chip images were applied to every frame via pixel-wise logical AND operations to isolate the activity of each lane. The number of white pixels was then counted in each corresponding ROI-applied snapshot, and the reduction in white pixels in each lane was taken as the area occupied by \textit{Physarum}, since \textit{Physarum} is black (pixel value=0) in the binary images. This occupied area was normalized by the total ROI white pixel count to yield the normalized area $X_{Vk}$. The custom Python scripts used for virtual mask generation and extraction of the lane state $X_{Vk}$ are available at \url{https://gitlab.com/Physarum_TSP/Image_processing_and_Lane_states_extraction/}.

\subsection{Estimation of power spectral density}
Welch’s method is a modified periodogram-based approach to estimate the power spectral density (PSD) of signals \cite{welch2003use}. This method reduces the noise in the data at the expense of frequency resolution. 
First, the time series $x(t)$ is divided into $n$ overlapping time segments of length $T$, each with an overlap of $d$ points. Each segment is then multiplied by a window function to reduce spectral leakage:
\begin{equation}
x_\alpha(t)=x(t+\alpha h) w(t), \hspace{10pt}  0 \leq t \leq T
\label{eq:window}
\end{equation}
In Equation \ref{eq:window}, $\alpha$ is the segment index, and $h$ represents the step size between consecutive segment starts. Next, the periodogram for each windowed segment is computed using the discrete Fourier transform:
\begin{equation}
P_\alpha(f)=\frac{1}{M}\left|\sum_{t=0}^{T-1} x_\alpha(t) e^{-j 2 \pi f t}\right|^2
\label{eq:periodogram}
\end{equation}
where $M$ is the normalization factor given by
$M=\sum_{t=0}^{T-1} w^2(t)$.
Finally, the PSD estimate is obtained by averaging the periodograms (as defined in Eq. \ref{eq:periodogram}) across all $n$ segments:
\begin{equation}
\operatorname{PSD}(f)=\frac{1}{n} \sum_{\alpha=0}^{n-1} P_\alpha(f)
\label{eq:PSD}
\end{equation}
Equation \ref{eq:PSD} provides the final estimate of the power spectral density for the entire signal.

In practice, PSDs were computed using the Welch method as implemented in the SciPy signal-processing module in Python, with each segment containing 256 data points (\texttt{nperseg =256}). Since the data were sampled at six-second intervals, the sampling frequency was $f_s=1 / 6\sim0.167 \mathrm{~Hz}$. PSDs were computed for each of the $N^2$ lanes and subsequently averaged for each Fourier frequency over the solution and non-solution groups within each of the pre-illumination, optical feedback, and NESS time intervals.

\subsection{Computation of problem size-aggregated means}
Consistent aggregation across trials for each problem size requires all normalized time-series and frequency-dependent quantities to be defined on a common grid. Since individual trials may have slightly different sampling points, direct pointwise averaging is not feasible. Each trial was therefore resampled onto a uniform grid using linear interpolation between adjacent data points for each grid point prior to aggregation, and the mean and standard error were computed pointwise across trials on this common grid. All aggregated quantities presented in the main text are computed using a grid of 200 points. For the synchronization index ratios $\overline{\delta}(t)$ and $\overline{\eta}(t)$, the analysis was also performed using a 500-point grid in addition to the standard 200-point grid; these results are shown in Fig. S11 of the Supplementary Material.

\textcolor{black}{\subsection{Statistical significance test}
Statistical significance was evaluated using paired $t$-tests. Differences in synchronization indices between the \textit{Physarum}-selected solution and alternative tours were considered statistically significant for $p<0.05$. Prior to performing the paired t-tests, the normality of the paired differences was assessed using the Shapiro-Wilk test. The statistical analyses described above were performed in Python using the SciPy library (scipy.stats).}

\subsection{Amoeba-inspired algorithms}


For simulations, we used AmoebaTSP, a computational toy model proposed to reproduce the linear scaling of computation time in the \textit{Physarum}-based TSP solving experiments, with the number of cities $N$ \cite{zhu2018remarkable}. In this model, intracellular resources necessary for branch growth are assumed to be supplied from the central processing disk at a constant rate. Both the inflow ($\Delta_{\text{in}}$) and outflow ($\Delta_{\text{out}}$) rates are set to 0.001.

Miyajima et al. \cite{miyajima2024proposed} proposed a modified AmoebaTSP model to reduce the number of iterations required for convergence to a feasible TSP solution \cite{miyajima2024proposed}.
One key modification involved replacing random uniform noise with random Gaussian noise, which significantly reduced the iteration count. Additionally, noise was incorporated directly into the branch (lane) state update rather than being applied to the sigmoidal activation function in Eq. \ref{eq:LVk}, resulting in tighter iteration bounds. Using the `Improved AmoebaTSP,' we conducted 1,000 trials for $N=10$ to $100$ and reproduced the $\sqrt{N}$ scaling of the iteration count as reported in \cite{miyajima2024proposed}. For comparison, we also performed 1,000 trials of the original AmoebaTSP for $N=10$ to $30$ cities. Results for the original AmoebaTSP were presented up to 30 cities only, as running this algorithm for more than 30 cities scaled inefficiently (see Section S4 of the Supplementary Material).

Each of these 1000 trials was executed in parallel at Oak Ridge Leadership Computing Facility on Frontier, one of the world's fastest exascale supercomputers with a total of 9,408 Advanced Micro Devices (AMD) compute nodes. Each Frontier compute node features a 64-core, AMD-optimized, third-generation EPYC CPU for high-performance applications, supporting two hardware threads per core and 512 GB DDR4 memory. 

\textcolor{white}{\cite{acharya2016application,agarwal2017eeg,kawala2020comparison,arora1996polynomial, idzenga2023experimental, reinelt1991tsplib, hopfield2025nobela, ventura1998quantum, rotondo2018open, MAASS2004593, 8697729, aono2012amoeba}}
 
\section*{Acknowledgments}
We acknowledge financial support from the Howard University (HU) Office of Research, the HU Graduate School, the IBM-HBCU Quantum Center at HU, and the Alfred P. Sloan Foundation's Matter-to-Life Program. This work utilized resources from the Leadership Computing Facility at Oak Ridge National Laboratory for running large TSP instances. Portions of this work were presented at the 2025 Quantum Biology Gordon Research Conference in Italy, the Sloan Foundation Matter-to-Life meetings in New York and Germany, the Active Solids Program at the Kavli Institute for Theoretical Physics (KITP) in California, the Quantum Thermodynamics Conference in Maryland, the Princeton-Texas A\&M-Casper Quantum Summer School in Wyoming, the Molecular Biophysics Workshop in France, and the Annual Meeting of the APS Division of Atomic, Molecular and Optical Physics in Oregon. The data analyzed in this paper were provided by one of the authors (MA) and Liping Zhu, who conducted experiments to collect the data at RIKEN prior to the publication of Refs. \cite{zhu2013amoeba, iwayama2016decision, zhu2018remarkable}. We would like to express our gratitude to these authors as well as to everyone involved in the related research. We also thank Prof. Masahito Mochizuki and Yusuke Miyajima for their insights and support regarding the Improved AmoebaTSP algorithm. KITP is supported in part by a grant from the National Science Foundation. 
\textcolor{black}{\subsection*{Data, Code, and Materials Availability}}
\textcolor{black}{All data and code needed to evaluate and reproduce the results in this paper are available through the GitLab repository linked in the Materials and Methods section. Representative sample data, including the original and color-coded videos for a representative trial, are available in the repository. The Python code used for data extraction and the corresponding extracted dataset are also available in the repository. Additional experimental data are available from the authors upon reasonable request. This study did not generate new materials. However, the \textit{Physarum} cross strain used in this study (Hokkaido University strain HU195$\times$200) can be requested from the authors.}

\newpage


\clearpage 

\bibliography{aapmsamp} 
\bibliographystyle{sciencemag}

\end{document}